\documentclass[%
11pt,
reprint,
onecolumn,
tightenlines,
superscriptaddress,
preprintnumbers,
nofootinbib,
amsmath,amssymb,amsthm,
aps,physrev,
floatfix,
eqsecnum,
]{revtex4-2}

\usepackage{isomath}
\usepackage{amsmath,amsthm}
\usepackage{amsbsy}
\usepackage{amssymb}
\usepackage{amscd}
\usepackage{amsfonts}
\usepackage{stmaryrd}
\usepackage{siunitx}
\usepackage{euscript}
\usepackage[utf8]{inputenc}
\usepackage[T1]{fontenc}
\usepackage{newtxtext} 
\everymath{\displaystyle}
\usepackage{exscale}
\usepackage{microtype}
\usepackage{hyperref}
\usepackage{booktabs}

\usepackage{graphicx}
\usepackage{boxedminipage}
\usepackage{calc}
\graphicspath{ {media/} }
\usepackage[caption=false,justification=raggedright]{subfig}

\usepackage{setspace}
\usepackage{enumitem}
\setitemize{noitemsep,topsep=0pt,parsep=0pt,partopsep=0pt}
\setenumerate{noitemsep,topsep=0pt,parsep=0pt,partopsep=0pt}
\setdescription{noitemsep,topsep=0pt,parsep=0pt,partopsep=0pt}

\usepackage{orcidlink}
\usepackage{siunitx}
\usepackage[small]{titlesec}

\titlespacing*{\section}{0pt}{12pt plus 4pt minus 2pt}{2pt plus 2pt minus 2pt}
\titlespacing*{\subsection}{0pt}{12pt plus 4pt minus 2pt}{2pt plus 2pt minus 2pt}
\titlespacing*{\subsubsection}{0pt}{12pt plus 4pt minus 2pt}{2pt plus 2pt minus 2pt}
\titlespacing*{\paragraph}{0pt}{12pt plus 4pt minus 2pt}{2pt plus 2pt minus 2pt}

\makeatletter

    \renewcommand*{\p@subsection}{}
    
    \renewcommand*{\p@subsubsection}{}
\makeatother

\usepackage{isomath}
\usepackage{amsmath}
\usepackage{amssymb}
\usepackage{amscd}
\usepackage{amsfonts}

\newcommand{\eps}{{\varepsilon}}

\newcommand{\half}{\frac{1}{2}}

\newcommand{\bfsigma}{\mathbold {\sigma}}

\DeclareMathOperator{\trace}{tr}

\newcommand{\dm}{\ \mathrm{d}}

\newcommand{\bfd}{{\mathbold d}}
\newcommand{\bfe}{{\mathbold e}}

\newcommand{\bfn}{{\mathbold n}}

\newcommand{\bfx}{{\mathbold x}}
\newcommand{\bfy}{{\mathbold y}}

\newcommand{\bfI}{{\mathbold I}}

\usepackage{enumitem}

\newcommand{\bfvareps}{\mathbold {\varepsilon}}

\begin{document}


\preprint{To appear in Journal of Applied Mechanics (\href{https://doi.org/10.1115/1.4068510}{DOI:10.1115/1.4068510}).}

\title{Crack Face Contact Modeling is Essential to Predict Crack-Parallel Stresses}

\author{Maryam Hakimzadeh}
    \email{mhakimz1@jhu.edu}
    \affiliation{Department of Civil and Systems Engineering, Johns Hopkins University}

\author{Noel Walkington}
    \affiliation{Center for Nonlinear Analysis, Department of Mathematical Sciences, Carnegie Mellon University}

\author{Carlos Mora-Corral}
    \affiliation{Departamento de Matemáticas, Universidad Autónoma de Madrid}
    \affiliation{Instituto de Ciencias Matemáticas, CSIC-UAM-UC3M-UCM}

\author{George Gazonas}
    \affiliation{DEVCOM Army Research Laboratory}

\author{Kaushik Dayal}
    \affiliation{Department of Civil and Environmental Engineering, Carnegie Mellon University}
    \affiliation{Center for Nonlinear Analysis, Department of Mathematical Sciences, Carnegie Mellon University}
    \affiliation{Department of Mechanical Engineering, Carnegie Mellon University}
    


\begin{abstract}
    Phase-field fracture models provide a powerful approach to modeling fracture, potentially enabling the unguided prediction of crack growth in complex patterns.
    To ensure that only tensile stresses and not compressive stresses drive crack growth, several models have been proposed that aim to distinguish between compressive and tensile loads.
    However, these models have a critical shortcoming: they do not account for the crack direction, and hence they cannot distinguish between crack-normal tensile stresses that drive crack growth and crack-parallel stresses that do not.

    In this study, we apply a phase-field fracture model, developed in our earlier work, that uses the crack direction to distinguish crack-parallel stresses from crack-normal stresses.
    This provides a transparent energetic formulation that drives cracks to grow in when crack faces open or slide past each other, while the cracks respond like the intact solid when the crack faces contact under normal compressive loads.
    We compare our approach against two widely used approaches, Spectral splitting and the Volumetric-Deviatoric splitting, and find that these predict unphysical crack growth and unphysical stress concentrations under loading conditions in which these should not occur.
    Specifically, we show that the splitting models predict spurious crack growth and stress concentration under pure crack-parallel normal stresses.
    However, our formulation, which resolves the crack-parallel stresses from the crack-normal stresses, predicts these correctly.
\end{abstract}

\maketitle

\section{Introduction}

Phase-field fracture models introduce a new field, in addition to the deformation, that indicates if a material element is intact or fractured, and then regularize this phase-field using higher-order derivatives.
The higher-order derivatives endow the phase-field with sufficient smoothness to enable the use of standard numerical methods to predict crack growth without specifying crack paths beforehand. 
That is, rather than sharp singular cracks as in the classical approach to fracture, cracks are now regularized to be nonsingular variations of the phase-field, making them
amenable to standard numerical approaches. 
The elastic response is coupled to the phase-field by setting that the regions that are fractured cannot carry any load, that is, the elastic energy density goes to zero in fractured regions.

However, the combination of crack regularization to a finite volume and setting the fractured elastic energy to zero causes unphysical behavior, such as crack growth under compressive loads.
To address these issues, \cite{miehe2010thermodynamically} proposed a partition of the elastic energy into tensile and compressive parts (denoted the ``Spectral'' split), and allowed the damaged region to sustain only the compressive part. 
An alternative class of approaches, following \cite{amor2009regularized}, proposed a partition of the energy into compressive hydrostatic, tensile hydrostatic, and
deviatoric parts (denoted the ``Volumetric-Deviatoric'' or ``VolDev'' split), and allowed the damaged region to sustain only the compressive hydrostatic part.
This has been extended to the finite deformation setting in \cite{tian2022mixed,xing2023adaptive,najmeddine2024efficient,tang2019phase}; however, none of these works consider the orientation of the crack.
Other modifications of the energy decompositions include \cite{vicentini2023energy} that proposed the star-convex model as a modification of the volumetric-deviatoric decomposition; \cite{wang2021phase} that extended the spectral decomposition to account for mixed mode fracture; and \cite{van2020strain} to account for anisotropy.
These approaches are reviewed and compared in \cite{zhang2022assessment,gupta2024damage,rahaman2022open,clayton2021nonlinear} and elsewhere.

However, all of the Spectral and VolDev splitting approaches have a key shortcoming: they do not account for the orientation of the crack.
Compressive stresses along the crack face and across the crack face are both treated in exactly the same way.
For instance, tension across the crack faces that drives crack growth is not distinguished from tension along the length of the crack, that does not directly drive crack growth but it is important in other ways.
For instance, \cite{bavzant2022critical} highlight that phase-field models do not accurately capture the influence of crack-parallel stresses, which significantly impact fracture energy and the fracture process zone width; they argue that the inability of phase-field models to incorporate the tensorial nature of stress interactions leads to discrepancies when compared to experimental results from the gap test.
Similarly, \cite{nguyen2020gap} emphasize that phase-field models lack the ability to explicitly account for the history-dependent effects of T-stress, which are crucial for realistic simulations of mixed-mode fracture in concrete and quasibrittle materials.

\begin{figure*}[htb!]
    \includegraphics[width=0.7\textwidth]{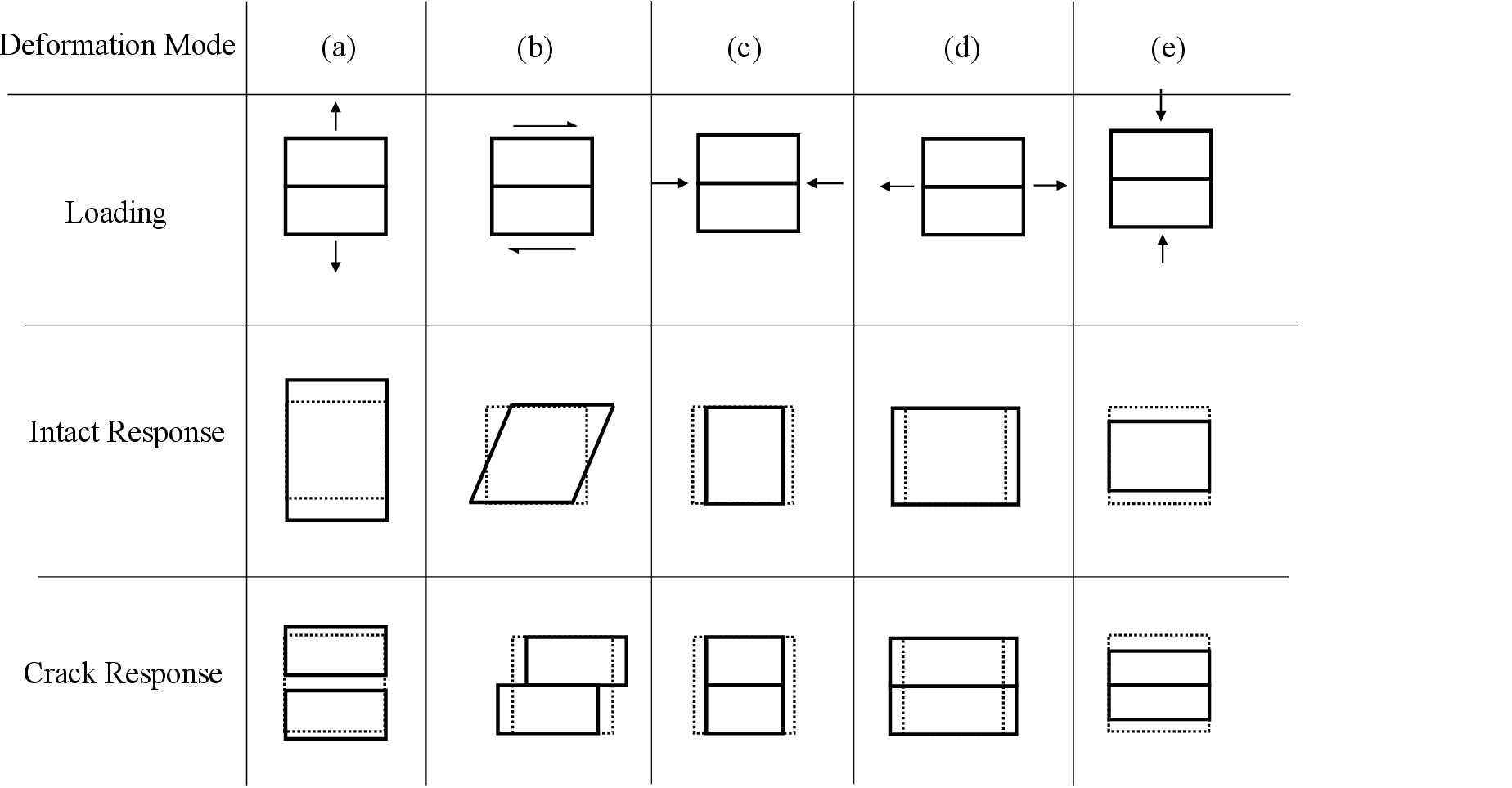}
    \caption{The top row shows different loadings, and the middle and lower rows show the idealized deformation for intact and cracked specimens, respectively. Based on this idealization, we assign zero energy to modes (a) and (b). The dotted lines in the second and third row show the undeformed configuration. This decomposition follows \cite{steinke2019phase,hakimzadeh2025phase}.}
    \label{fig:deform modes}
\end{figure*}

An important departure from the idea of splitting approaches is the work by \cite{steinke2019phase}, which introduced the idea that the crack orientation plays an essential role in determining the response.
Based on this idea, \cite{steinke2019phase,fei2020phase-cmame,fei2020phase-ijnme,agrawal2016multiscale} proposed modifications of the stress tensor to account for the difference in crack-normal and crack-parallel tractions.
While our approach in \cite{hakimzadeh2022phase,hakimzadeh2025phase} builds on the ideas of \cite{steinke2019phase}, an important distinction is that we modify the energy rather than the stress, and hence our model is hyperelastic.
In contrast, as shown in \cite{hakimzadeh2022phase}, the stress-based approach is not generally hyperelastic.
Our approach enables us to distinguish between crack-parallel stresses and crack-closing stresses as well as allow for crack faces sliding past each other.
Our approach applies a QR decomposition of the deformation gradient in the basis associated with the local crack normal, enabling us to distinguish between crack faces sliding past each other -- which is formulated to cost no energy; crack opening -- which is also formulated to cost no energy; and crack closing and crack face contact -- which are formulated to have an elastic response that is identical to the intact material.
This is shown schematically in Figure \ref{fig:deform modes}.

In this paper, we compare the phase-field fracture model developed in \cite{hakimzadeh2022phase,hakimzadeh2025phase}, which distinguishes between crack-parallel stresses and crack-closing stresses, against the Spectral \cite{miehe2010thermodynamically} and VolDev \cite{amor2009regularized} splitting approaches.
Specifically, in Section \ref{sec:T-stress}, we examine the effect of tensile and compressive crack-parallel stresses, and show that the crack growth and stresses predicted by the Spectral and VolDev approaches are unphysical, in contrast to the model from \cite{hakimzadeh2022phase}.


\section{Formulation of the Model} \label{sec:methodology}

\subsection{Classical Phase-Field Fracture Model}

The variational fracture model by Francfort and Marigo \cite{francfort1998revisiting} describes brittle fracture through energy minimization:
\begin{equation}\label{eq:energy_variational}
    E[\bfy, \Gamma] = \int_{\Omega \setminus \Gamma} W(\nabla \bfy) \,\mathrm{d}\Omega + G_c \mathcal{H}^{n-1}(\Gamma) + \text{work due to external loads},
\end{equation}
where $\bfy$ is the displacement, $\Gamma$ is the crack surface, $\Omega$ is the reference domain, $W$ is the elastic energy density, $G_c$ is the work to fracture, and $\mathcal{H}^{n-1}(\Gamma)$  is the Hausdorff area of $\Gamma$.
The model asserts that the deformation and the crack surface are obtained by minimizing $E$ subject to irreversibility, i.e., the crack cannot heal.

Due to the computational challenges in working with this model directly --- the solution lives in function spaces that are very challenging to approximate numerically --- a regularized phase-field formulation was introduced by Ambrosio and Tortorelli \cite{AmTo90,AmTo92}, inspired by \cite{MoMo77,Modica87}:
\begin{equation}\label{eq:energy_phase_field}
    E[\bfy, \phi] = \int_{\Omega} \left( \phi^2 W (\nabla \bfy) + G_c \left( \frac{(1-\phi)^2}{4\epsilon} + \epsilon |\nabla\phi|^2 \right) \right) \,\mathrm{d}\Omega + \text{work due to external loads},
\end{equation}
where $\phi(\bfx)$ is a scalar phase field that quantifies the level of damage at every referential point $\bfx$, and is sufficiently smooth to allow standard FEM approximation.
Specifically, $\phi$, bounded by $0 \leq \phi \leq 1$, indicates that the material is intact at $\phi(\bfx) \simeq 1$ and completely fractured at $\phi(\bfx) \simeq 0$.
The small parameter $\epsilon$ controls the crack width.
The method has been used very widely for fracture calculations and we do not review this extensive literature.

\subsection{Accounting for Crack-Face Contact}

A key limitation of the standard phase-field fracture model is its inability to distinguish between tensile, shear, and compressive loadings across or along the crack face, often leading to unphysical crack propagation under compression. 
Various modifications have been proposed to address this issue, but they fail to incorporate the crack orientation explicitly \cite{miehe2010thermodynamically, amor2009regularized, de2022nucleation, vicentini2023energy, tang2019phase, van2020strain, wang2021phase, zhang2022assessment}. 
As noted in \cite{steinke2019phase} and further demonstrated in \cite{hakimzadeh2022phase, hakimzadeh2025phase} and elsewhere, correctly capturing crack behavior under mixed-mode loading requires a formulation that accounts for the crack-face normal.

To overcome this limitation, we adopt the phase-field formulation introduced in \cite{hakimzadeh2022phase}, which incorporates an effective crack energy density $W_d$ that depends on the crack orientation. 
Using the QR (or Gram-Schmidt) decomposition of the deformation gradient $\nabla\bfy$, e.g. \cite{clayton2020constitutive}, in the basis of the crack enables us to distinguish the various modes shown in Figure \ref{fig:deform modes}.
We then assigned zero energy to modes in which the crack faces slide past each other or open up, and assign the intact response to modes in which the crack faces contact.
This ensures a physically consistent response, particularly under compressive loading, by preventing unphysical interpenetration of crack faces, as demonstrated through several examples in \cite{hakimzadeh2022phase, hakimzadeh2025phase}.

Our approach uses a free energy of the form
\begin{equation}
\label{eqn:model}
    E[\bfy,  \phi]  
    = 
    \int_{\Omega} \left( \phi^2 W (\nabla \bfy) + \left( 1-\phi^2 \right)  W_d (\nabla \bfy,\bfd) \right)  \dm\Omega + G_c \int_{\Omega} \left( \frac{(1-\phi)^2}{2\epsilon} + \frac{\epsilon}{2}  |\nabla\phi|^2 \right) \dm\Omega  + \text{ work due to external loads. }
\end{equation}
The key feature of this model is the introduction of the effective crack energy density $W_d$ that is active in the fractured volumes.
It imparts the properties of an idealized sharp crack to the regularized (finite volume) phase-field crack. 
The function $W_d$ takes as argument the deformation gradient, as well as a vector field $\bfd$ that corresponds to the normal to the crack face which is essential to be able to decompose the deformation into the modes shown in Figure \ref{fig:deform modes}.
However, $\bfd$ is not an independent field; rather it is a function of $\phi$ as described in \cite{hakimzadeh2025phase}.
For brevity, we do not present the detailed expressions for $W_d$ but refer instead to \cite{hakimzadeh2022phase, hakimzadeh2025phase}.
In this paper, we use an isotropic linear elastic response, with Lam\'e moduli $\lambda$ and $\mu$.

\subsection{Energy Splitting Approaches}

A simplistic approach to modeling the energy density of damaged volumes is to assume that the damaged region carries zero elastic energy. 
However, this assumption often leads to unphysical behaviors, such as the interpenetration of crack faces, incorrect mechanical response when the crack closes, and crack growth under compression. 
To address these shortcomings, alternative formulations have been proposed, notably by the Spectral split by \cite{miehe2010thermodynamically} and the VolDev split by \cite{amor2009regularized}, which decompose the energy and selectively associate specific energy components with the damaged region. 
While these methods offer improvements over the zero-energy assumption, they do not account for crack orientation in the energetic decomposition, leading to unphysical predictions under complex stress states.

\subsubsection{Spectral Splitting} 

Following \cite{miehe2010thermodynamically}, the spectral splitting approach defines the compressive ($\psi_0^-$) and tensile ($\psi_0^+$) energies as
\begin{equation}
    \psi_0^{\pm} 
        := 
        \half\lambda\langle\eps_1 + \eps_2 +\eps_3 \rangle_\pm^2
        + \mu \left( \langle\eps_1\rangle_\pm^2 + \langle\eps_2\rangle_\pm^2 + \langle\eps_3\rangle_\pm^2 \right) ,
\end{equation}
where $\eps_1, \eps_2, \eps_3$ are the principal strains with the corresponding principal directions $\bfn_1, \bfn_2, \bfn_3$. 
Here, $\langle x \rangle_{+} := \max \{0, x\}$ and $\langle x \rangle_{-} := \min \{0, x\}$, and only the compressive energy $\psi_0^-$ is taken to contribute to the energy in the cracked region. 
This results in the stress
\begin{equation}
    \bfsigma_{split} 
    = 
        \sum_{a=1}^3 
            \left(\lambda\langle\eps_1+\eps_2+\eps_3\rangle_{-} + 2\mu\langle\eps_a\rangle_{-}\right)
            \bfn_a\otimes\bfn_a ,
\end{equation}
which inherently prevents tensile stresses in all directions.

In the Appendix of \cite{hakimzadeh2022phase}, we identified significant challenges associated with Spectral splitting. 
Specifically, this approach does not accurately capture crack response under complex loading conditions, such as uniaxial tension parallel to the crack . 
Despite these limitations, many studies on fracture modeling continue to rely on this approach or variants like those proposed by \cite{lo2019phase}, which builds on the work of \cite{miehe2010thermodynamically}. 
Recent studies in studying crack behavior under complex loadings, including \cite{sun2021poro}, utilize this extended method by \cite{lo2019phase}. 
Other examples of fracture modeling include recent work by \cite{clayton2022stress}, which adopts the formulation introduced by \cite{miehe2015phase}, and \cite{stocek2023viscoelastic}, which is built on the frameworks of \cite{miehe2017phase} and \cite{miehe2010thermodynamically}.

\subsubsection{Volumetric-Deviatoric Splitting} 

The VolDev split proposed by \cite{amor2009regularized} aims to allow cracks to resist compressive hydrostatic stresses while excluding tensile hydrostatic and deviatoric stresses. The energy is expressed as
\begin{equation}
    W(\bfvareps) = \half\kappa\langle\trace\bfvareps\rangle_-^2 + \half\kappa\langle\trace\bfvareps\rangle_+^2 + \mu|\bfvareps_D|^2,
\end{equation}
where $\kappa = \lambda+2\mu/3$ is the bulk modulus, and $\bfvareps_D := \bfvareps - \frac{1}{3}\left(\trace{\bfvareps}\right)\bfI$ is the deviatoric strain component. Here, only the compressive hydrostatic term $\half\kappa\langle\trace\bfvareps\rangle_-^2$ is retained for the cracked region's energy, giving the stress response
\begin{equation*}
    \bfsigma_{split} = \left(\kappa\langle\trace\bfvareps\rangle_-\right)\bfI,
\end{equation*}
producing a purely compressive hydrostatic response.

As discussed in the appendix of our previous work \cite{hakimzadeh2022phase}, the regularized approach proposed by \cite{amor2009regularized} encounters limitations in accurately capturing the crack response under complex stress states. These include uniaxial tension parallel to the crack plane and uniaxial compression normal to the crack. Despite these limitations, this approach remains widely used in several studies that model crack behavior in complex loadings.
\subsection{Numerical Implementation}

We employ the finite element method (FEM) for numerical simulations, using the open-source FEniCS library, which has been extensively utilized in mechanics applications \cite{logg2012automated, barchiesi2021computation}.
The details are provided in \cite{hirshikesh2019fenics} for the Spectral split approach, \cite{natarajan2019phase} for the VolDev split approach, and \cite{hakimzadeh2022phase} for our approach.

\section{Crack Growth and Stress Distribution under Crack-Parallel Stresses} 
\label{sec:T-stress}

Here, we analyze crack growth and the stress distribution when the crack is subject to crack-parallel (tensile and compressive) stresses using Spectral splitting, VolDev splitting, and our approach.

Consider a specimen containing a crack with normal $\bfe_2$, subjected to a far-field stress of $\sigma_0 \bfe_1 \otimes \bfe_1$, with $\sigma_0 > 0$. 
Assuming isotropic elasticity, this corresponds equivalently to the following far-field strain:
\begin{equation}
\label{eqn:split-ex-1}
    \bfvareps = \frac{\sigma_0}{\mu(3\lambda+2\mu)}
        \begin{pmatrix}
            \lambda+\mu & 0 & 0
            \\
            0 & -\lambda/2 & 0
            \\
            0 & 0 & -\lambda/2
        \end{pmatrix} .
\end{equation}

The resulting stress response within the crack using the Spectral splitting model is
\begin{equation}
    \bfsigma_{split} = -\frac{\sigma_0}{3\lambda+2\mu} \left(\bfe_2\otimes\bfe_2+\bfe_3\otimes\bfe_3\right) .
\end{equation}
We highlight two key unphysical aspects of this result.  
First, when the Poisson ratio $\nu$ is positive, we find an unphysical compressive stress in the $\bfe_2$ direction.  
Second, the stress along the $\bfe_1$ direction is zero, whereas under a far-field stress parallel to the crack, the material is expected to behave as an intact material, sustaining the far-field stress as it is.  
This incorrect stress distribution leads to spurious crack growth. 
The artificial lack of stiffness in tension parallel to the crack creates spurious stress concentrations near the crack tip, as the intact material ahead of the tip sustains the far-field stress in the $\bfe_1$ direction, while the cracked region does not. 
Moreover, this discrepancy can significantly affect the calculation of the T-stress \cite{bavzant2022critical}.

In the VolDev model, we obtain $\bfsigma_{split} = {\bf 0}$.  
However, the material should fully sustain the far-field tensile stress parallel to the crack.  
This unphysical response leads to spurious stress concentrations at the crack tip, which influence both the crack driving force and the calculation of the T-stress.

\subsection{Numerical Evaluation of Crack Growth and Stress Distributions}

We next numerically evaluate the crack path and stress distribution in a finite domain.
For a clear comparison, we use the same domain, mesh, and material properties across all three approaches, and first set $\nu = 0$ to eliminate crack faces separating or contacting due to the crack-parallel loading.
We then examine having $\nu>0$ and $\nu<0$.

In our calculations, we set the vertical displacement and the horizontal tractions to zero on the top and bottom faces; the horizontal displacement is controlled on the left and right faces; and the vertical tractions are set to zero on the left and right faces.

\paragraph{Crack-Parallel Tension.}

\begin{figure}[htb!]
    \subfloat[Initial crack, shown by $\phi$.]{
        \includegraphics[width=0.3\textwidth]{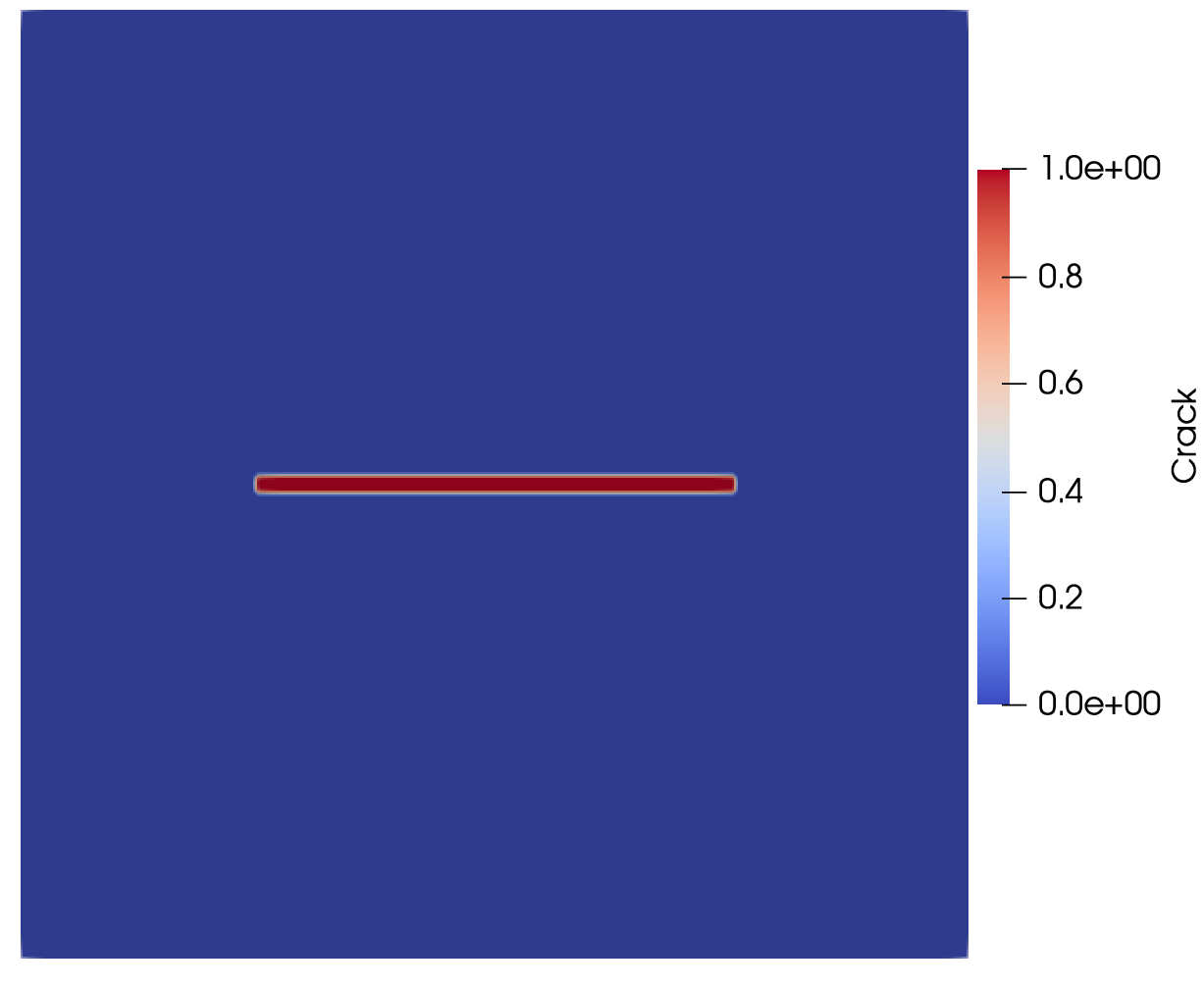}
    }
    \\
    \subfloat[No crack growth observed in our model.]{
        \includegraphics[width=0.35\textwidth]{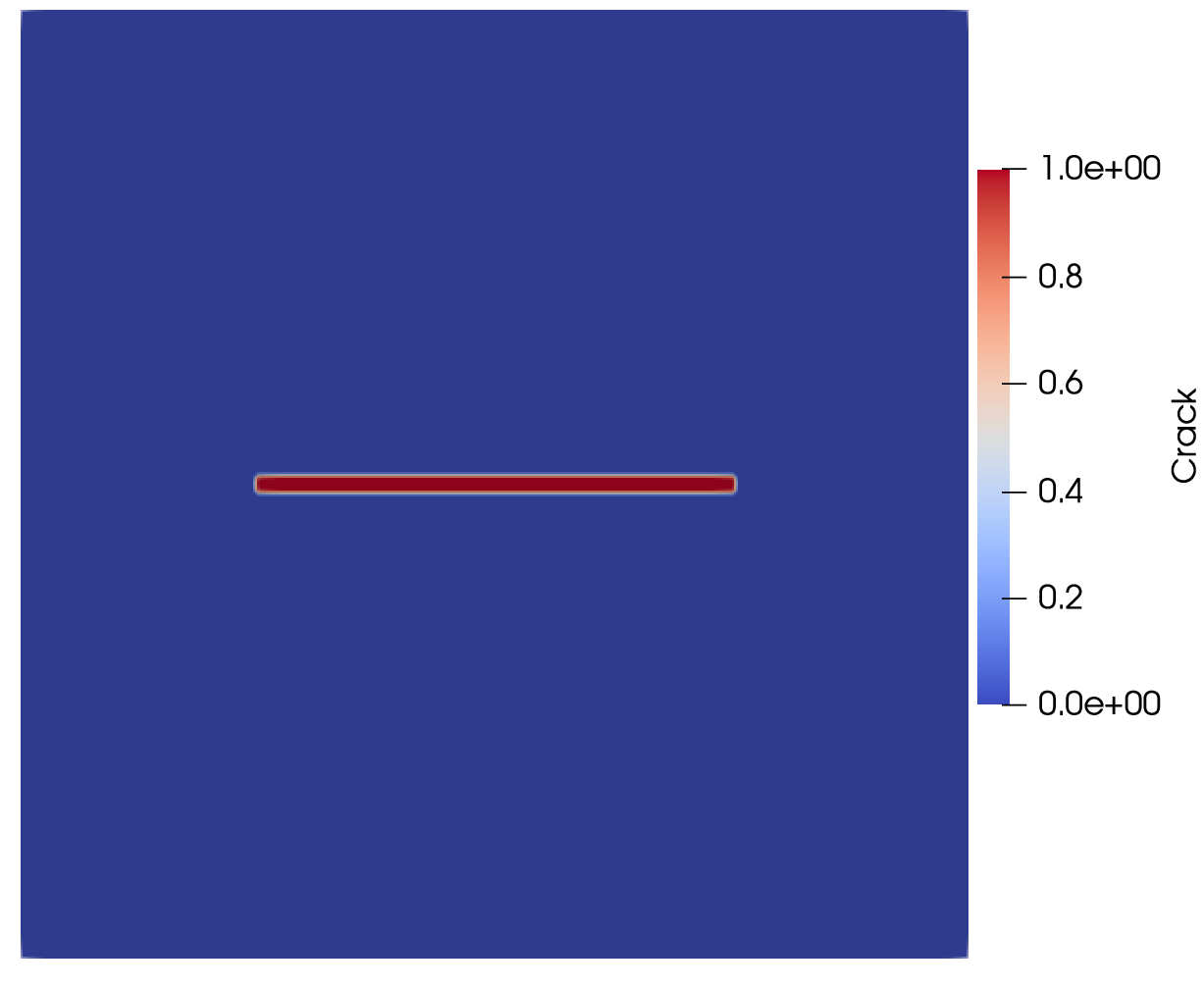}
    }
    \hspace{0.2\textwidth}
    \subfloat[Uniform stress distribution in our model.]{
        \includegraphics[width=0.35\textwidth]{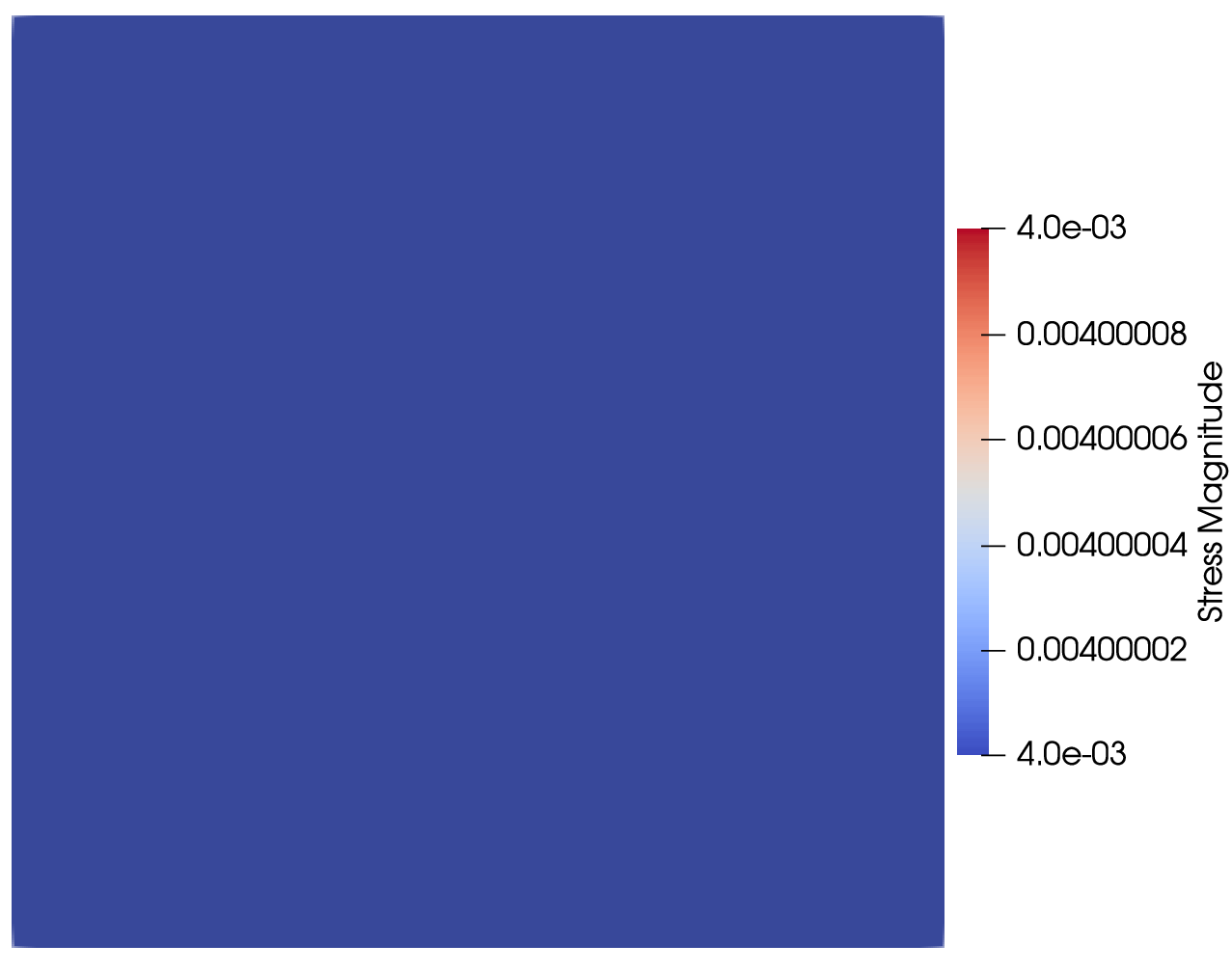}
    }
    \\
    \subfloat[Unphysical crack growth with Spectral splitting.]{
        \includegraphics[width=0.35\textwidth]{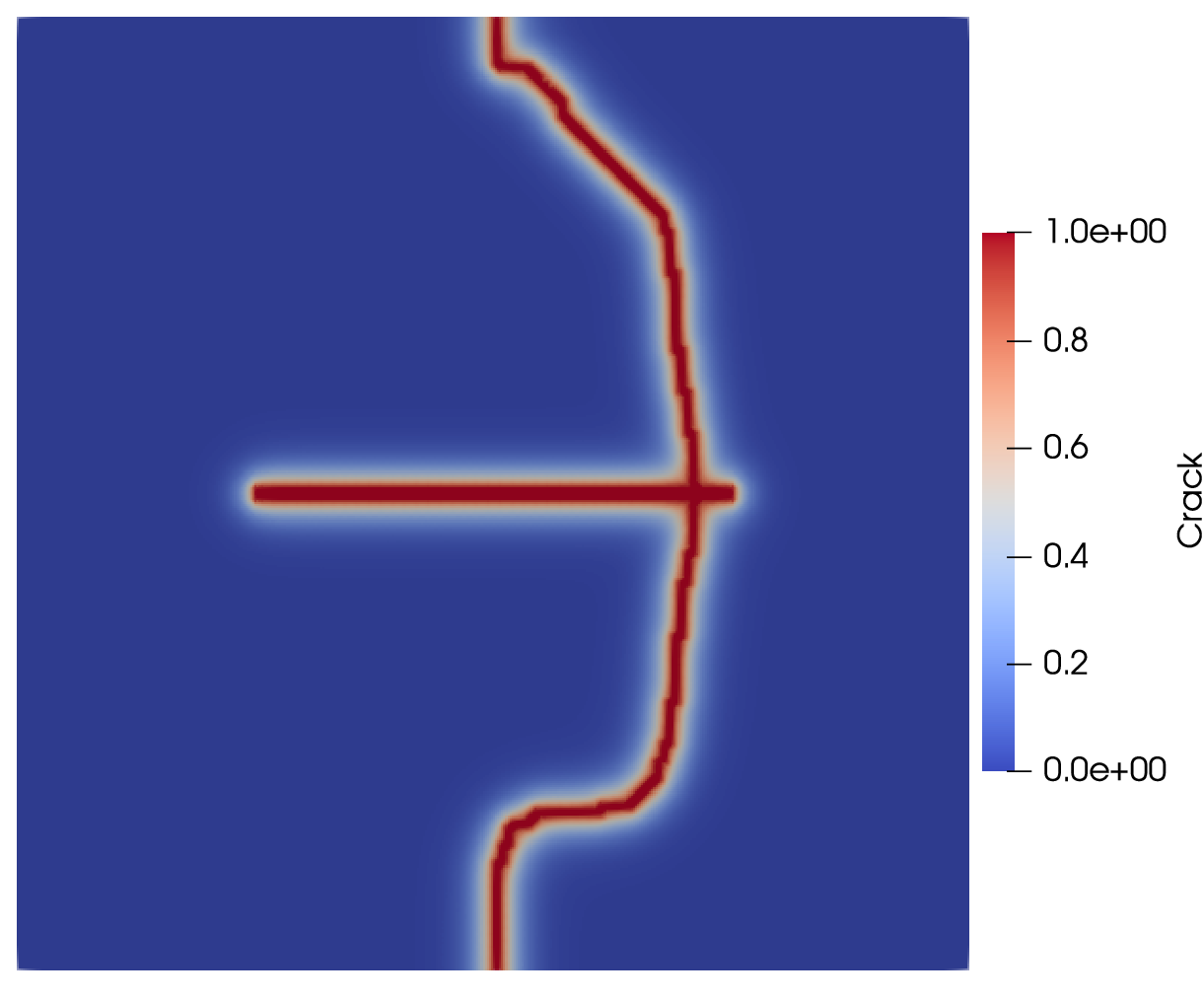}
    }
    \hspace{0.2\textwidth}
    \subfloat[Spurious stress concentrations with Spectral splitting.]{
        \includegraphics[width=0.35\textwidth]{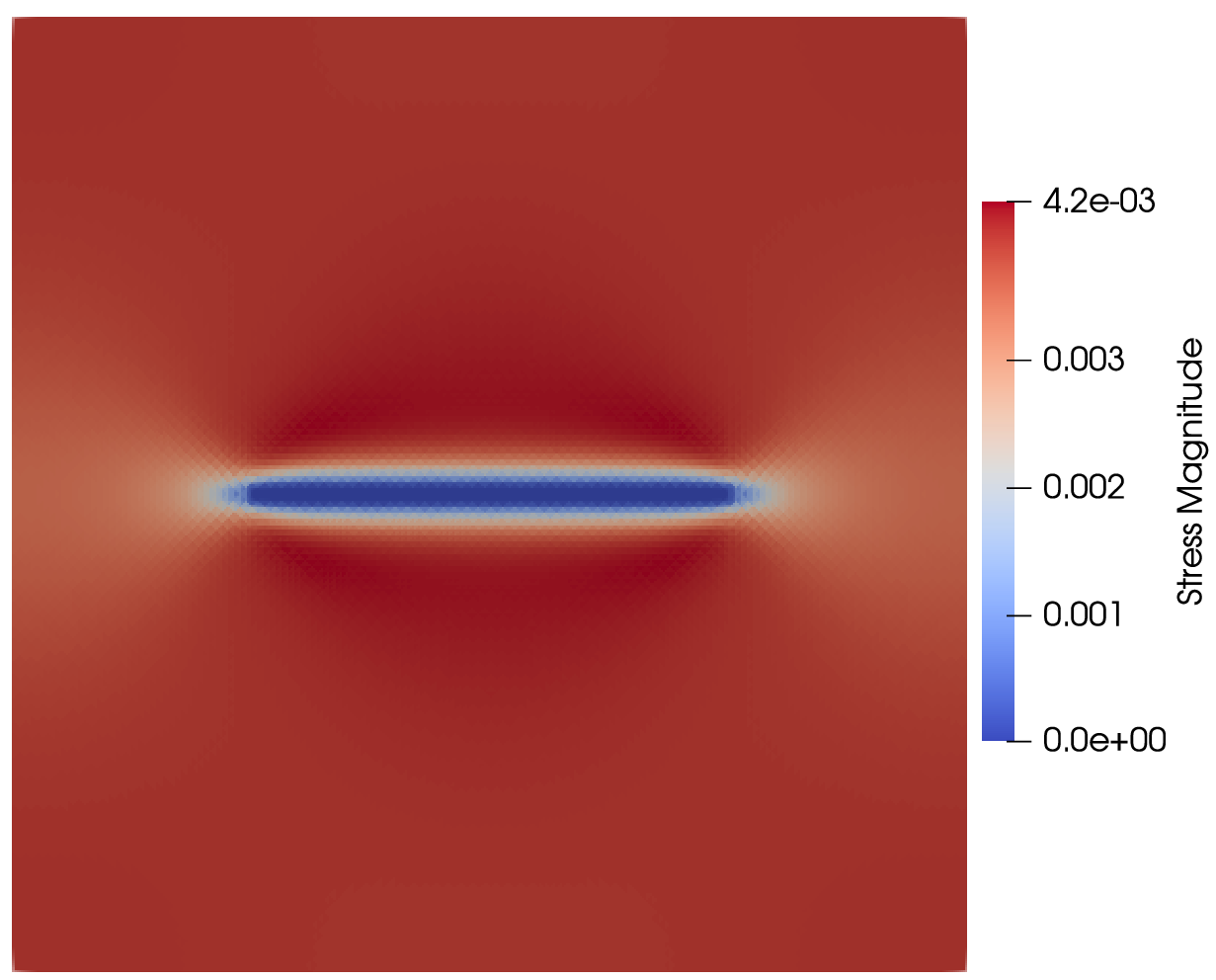}
    }
    \\    
    \subfloat[Unphysical crack growth with VolDev splitting.]{
        \includegraphics[width=0.35\textwidth]{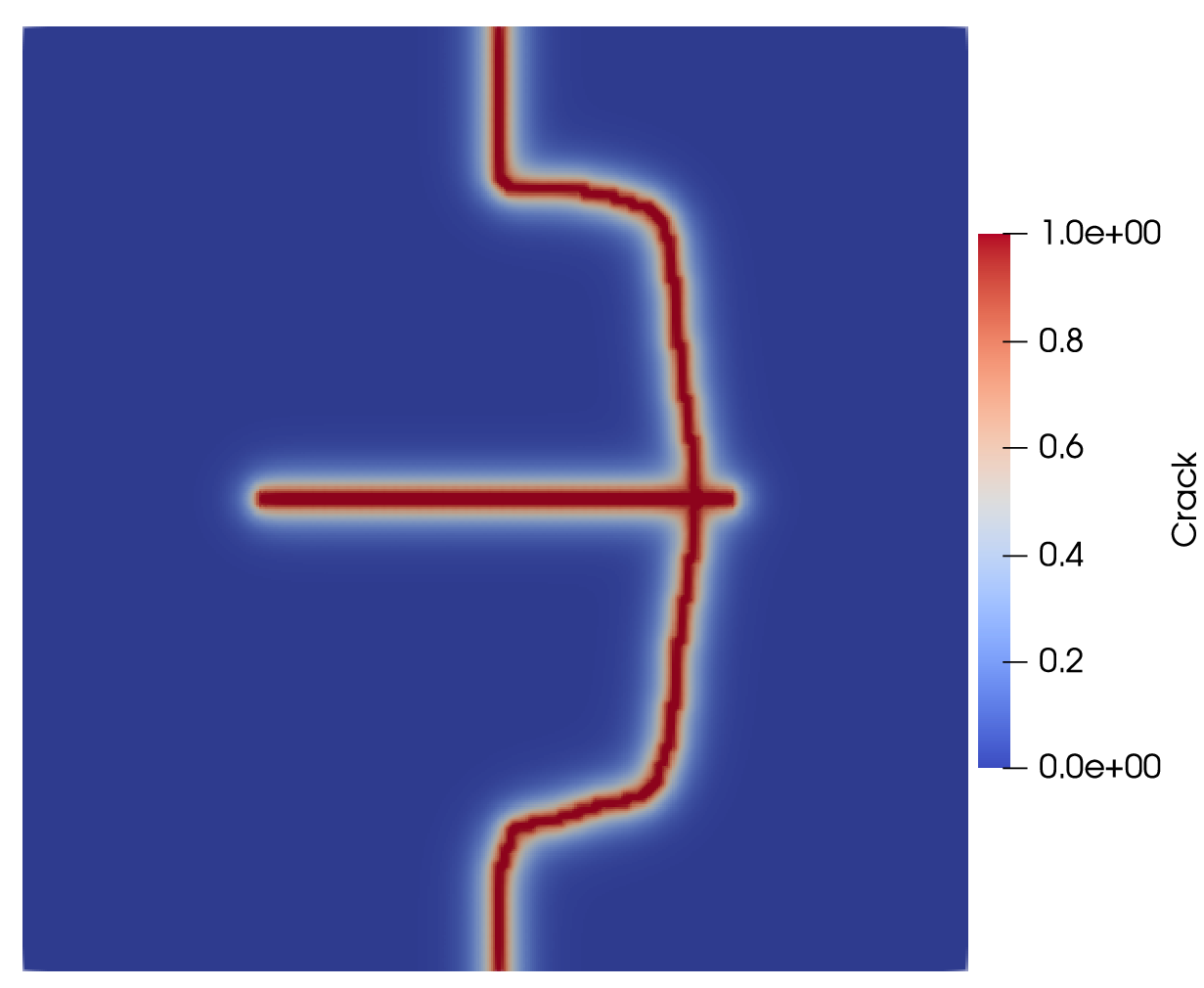}
    }
    \hspace{0.2\textwidth}
    \subfloat[Spurious stress concentrations with VolDev splitting.]{
        \includegraphics[width=0.35\textwidth]{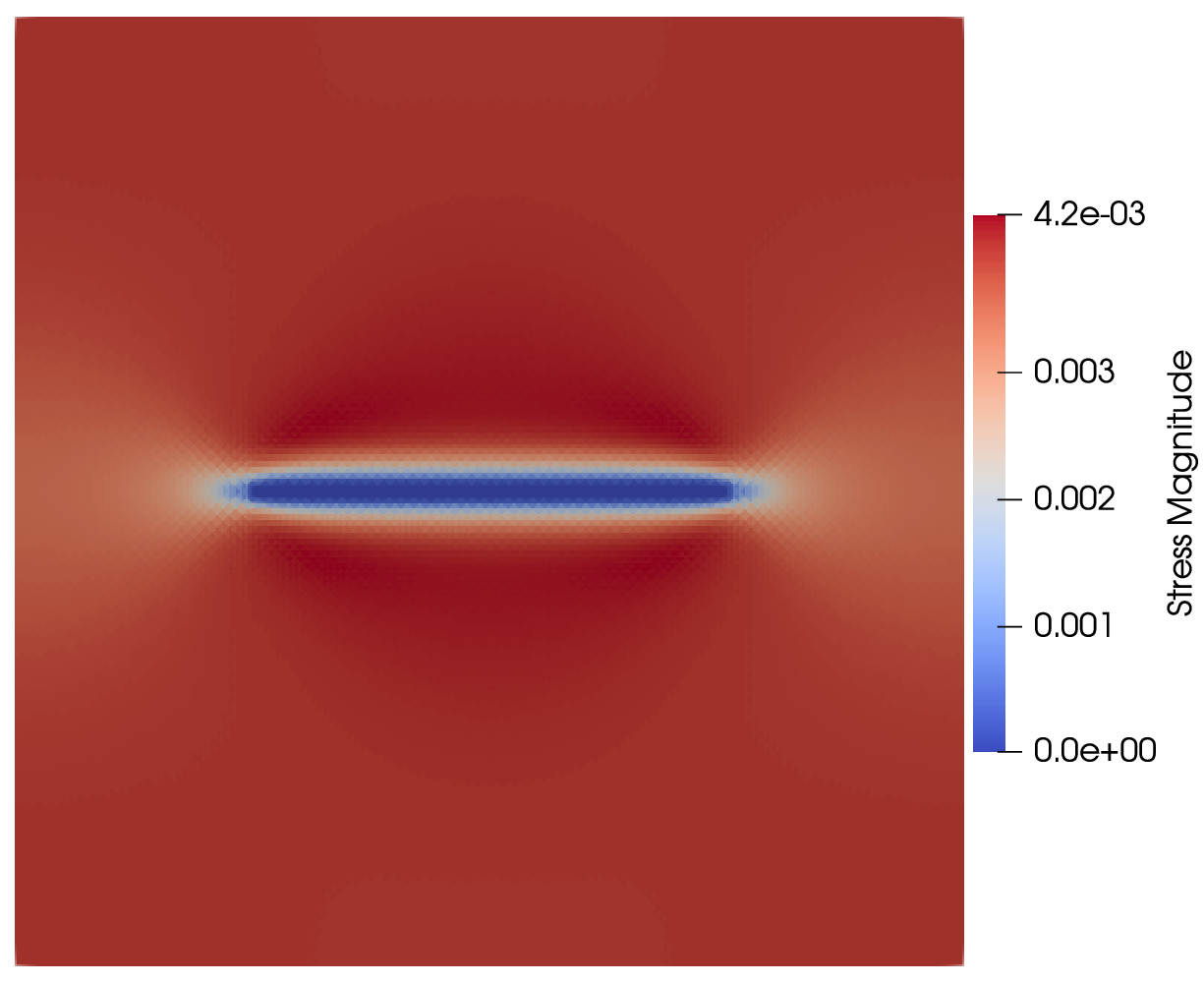}
    }
    \caption{Comparison of crack growth paths and stress fields under tensile crack-parallel stress.}
    \label{fig:T-stress_compare}
\end{figure}

We first examine the effect of tensile loading in Figure \ref{fig:T-stress_compare}, which shows the initial crack, the subsequent crack growth, and the stress distribution at the first load step, for all 3 models.
For simplicity of interpretation, we set $\nu = 0$.
We observe that our approach predicts essentially a uniform stress distribution, that corresponds to the cracked specimen having the same response as an intact specimen, and there is no crack growth.
Both of these features correspond to physical expectations.

In both the Spectral and VolDev splits, the stresses are significantly non-uniform with spurious stress concentrations at the crack tips that were anticipated by the simple analysis above.
The spurious stress distribution, in turn, drives unphysical crack growth.

\paragraph{Crack-Parallel Compression.}

\begin{figure}[htb!]
    \centering
    \subfloat[Initial crack, shown by $\phi$.]{
        \includegraphics[width=0.3\textwidth]{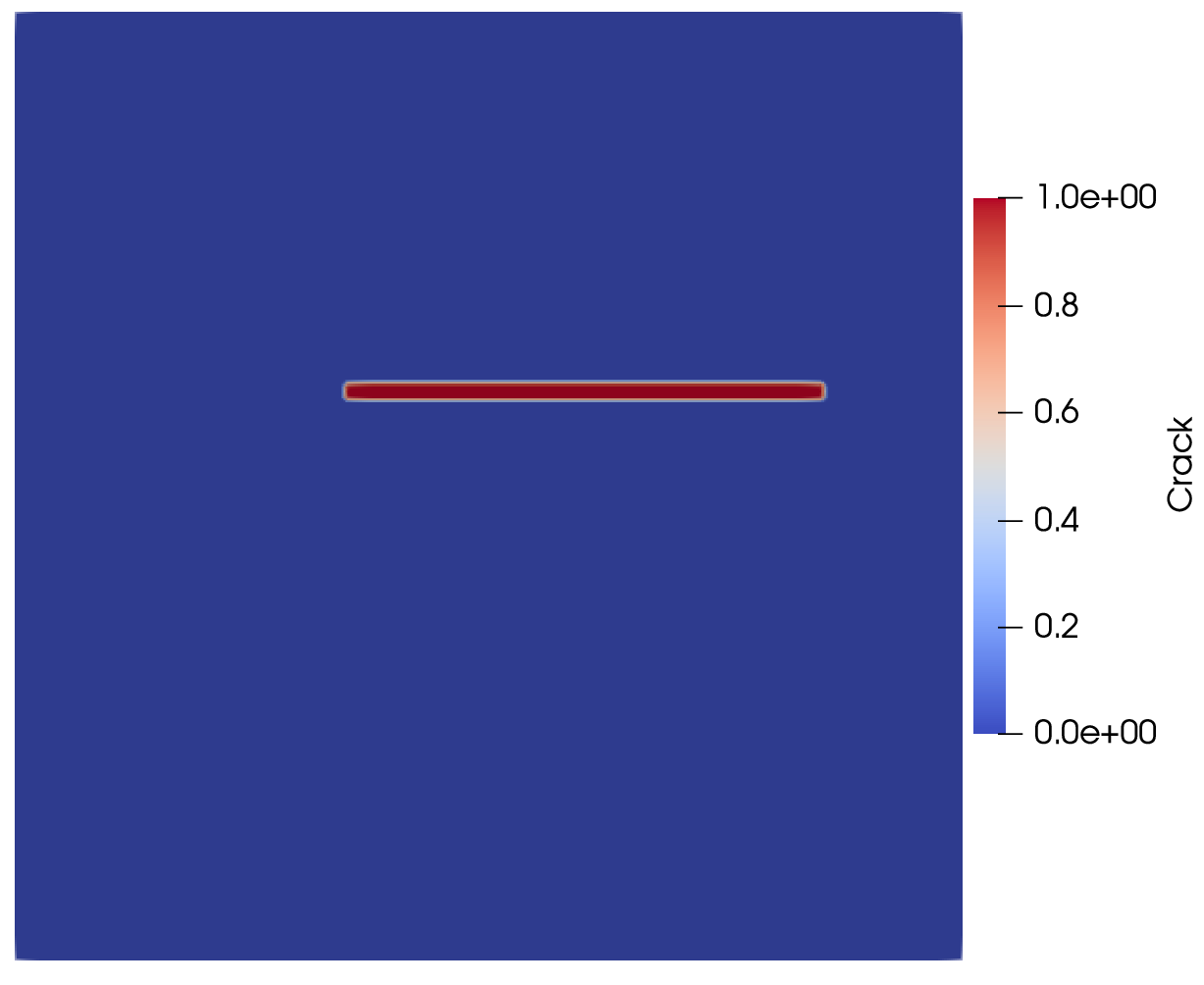}
    }
    \\
    \subfloat[No crack growth observed in our model.]{
        \includegraphics[width=0.35\textwidth]{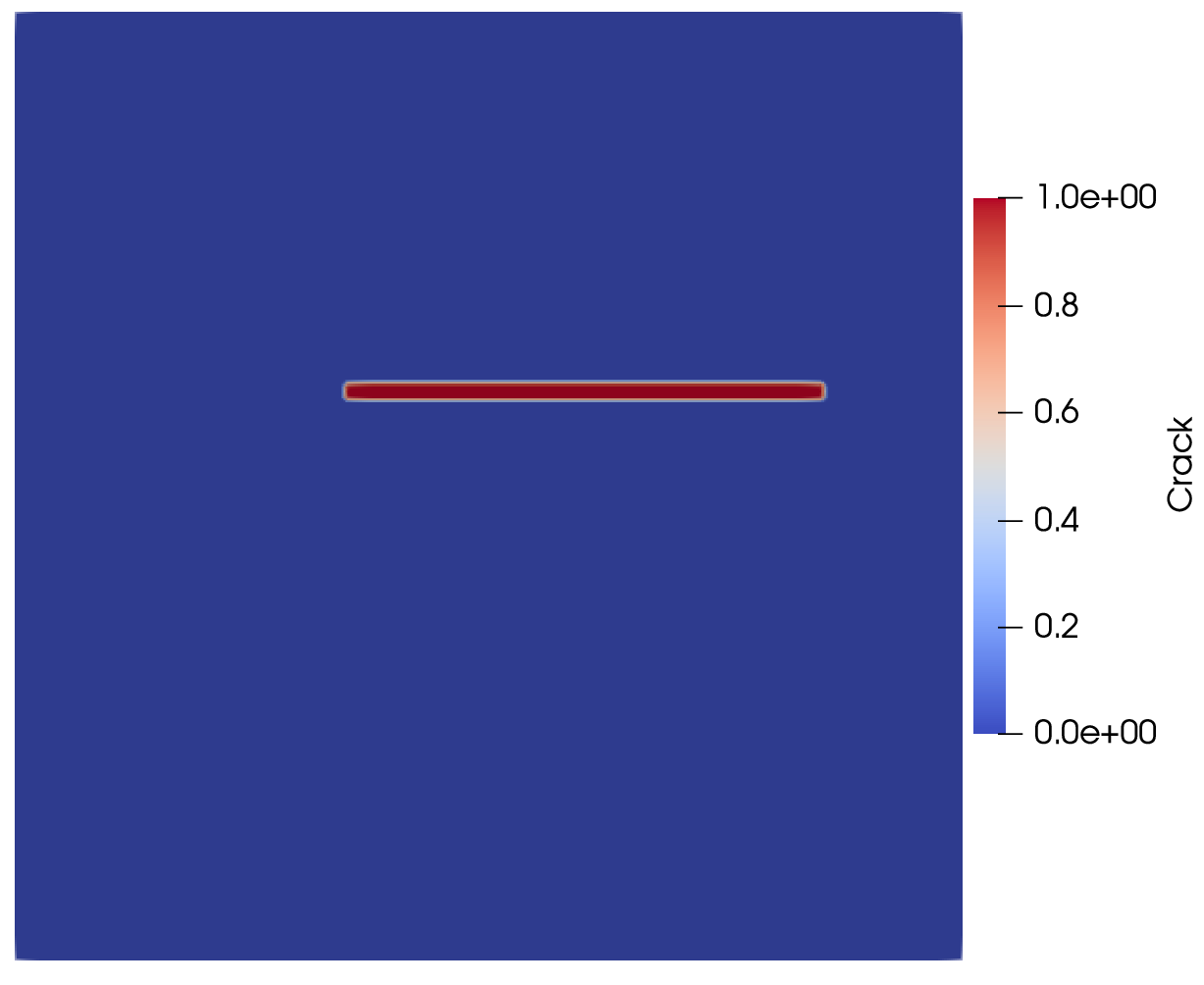}
    }
    \hspace{0.2\textwidth}
    \subfloat[Uniform stress distribution in our model.]{
        \includegraphics[width=0.35\textwidth]{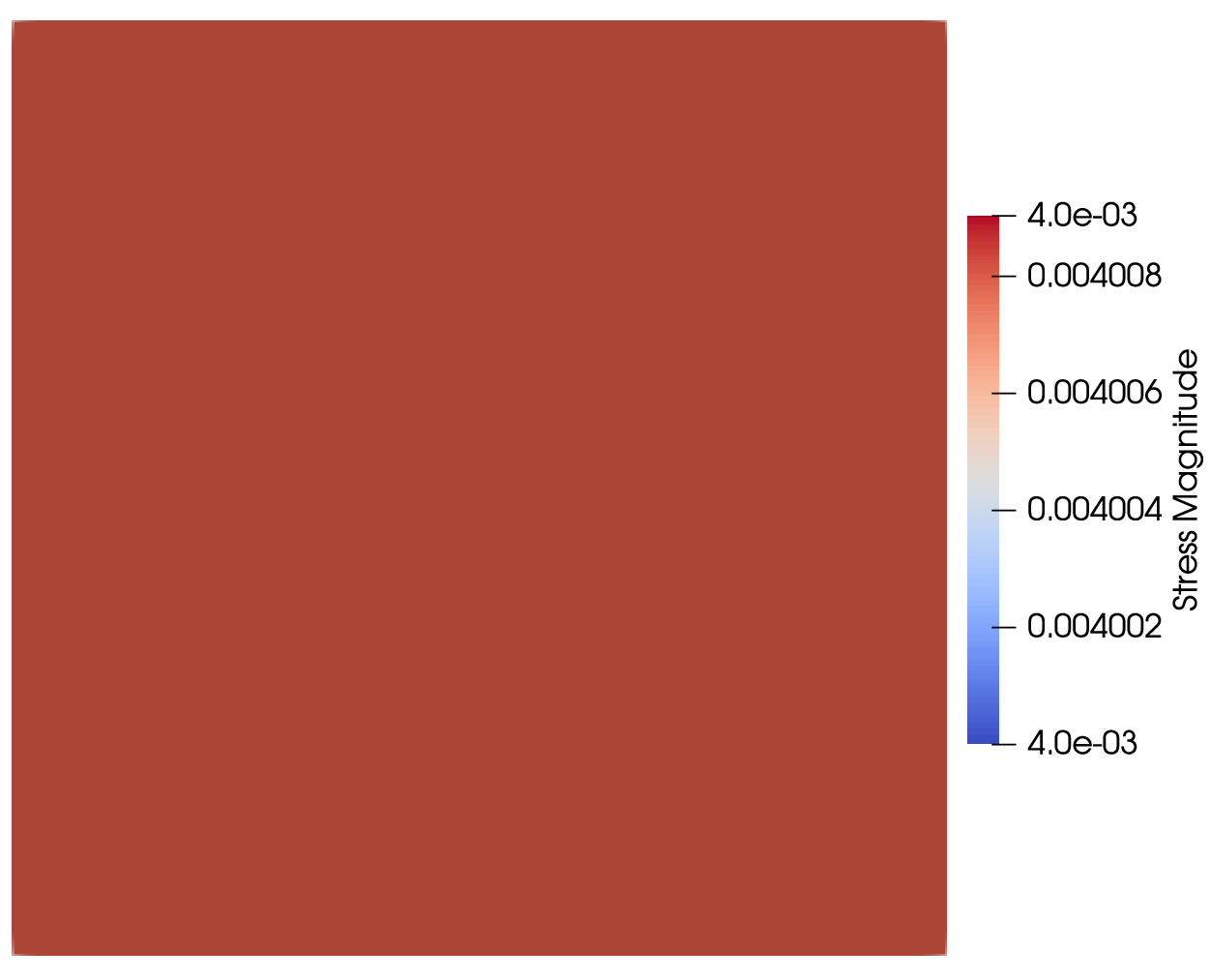}
    }
    \\
    \subfloat[No crack growth with Spectral splitting.]{
        \includegraphics[width=0.35\textwidth]{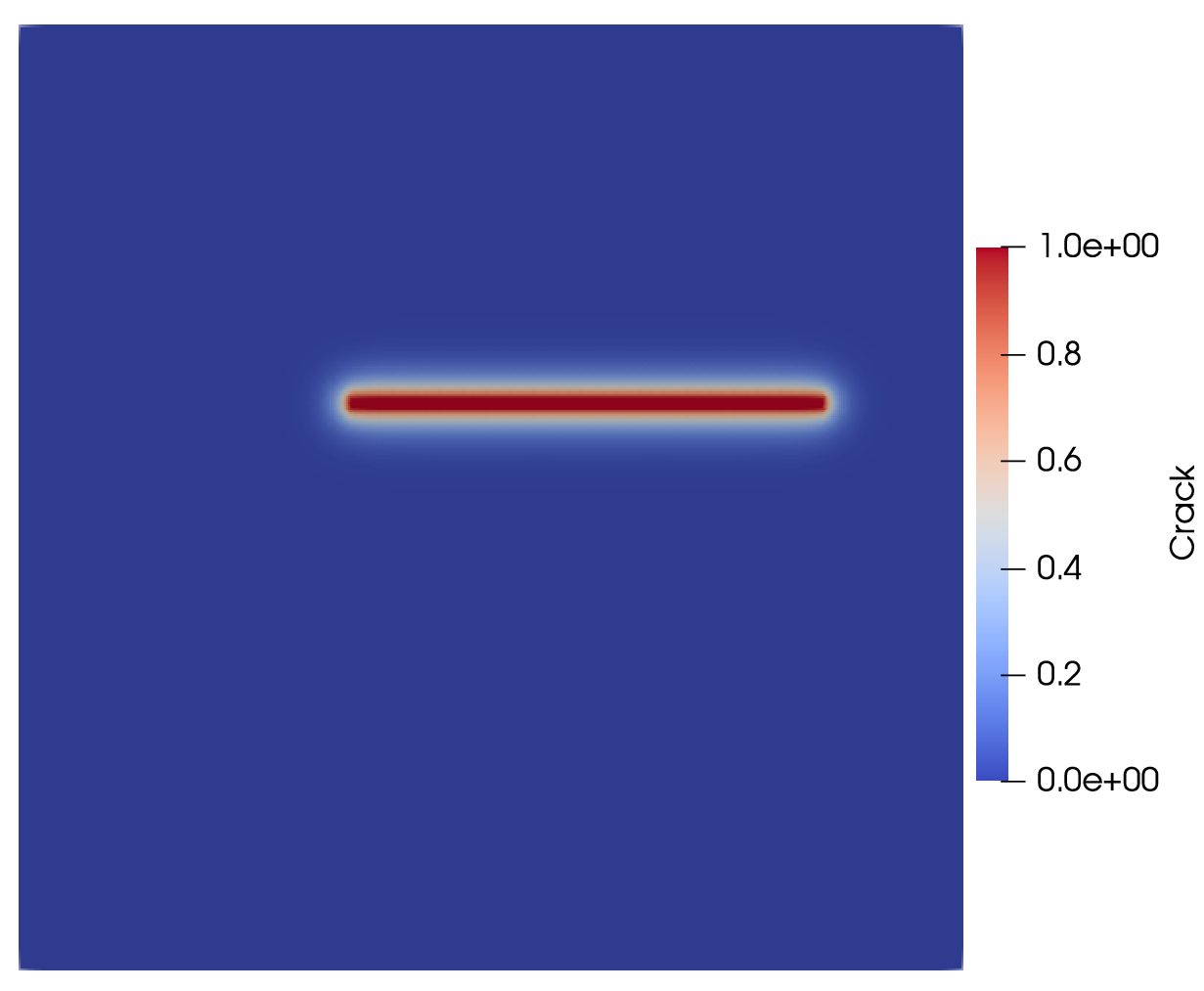}
    }
    \hspace{0.2\textwidth}
    \subfloat[Spurious stress concentrations with Spectral splitting.]{
        \includegraphics[width=0.35\textwidth]{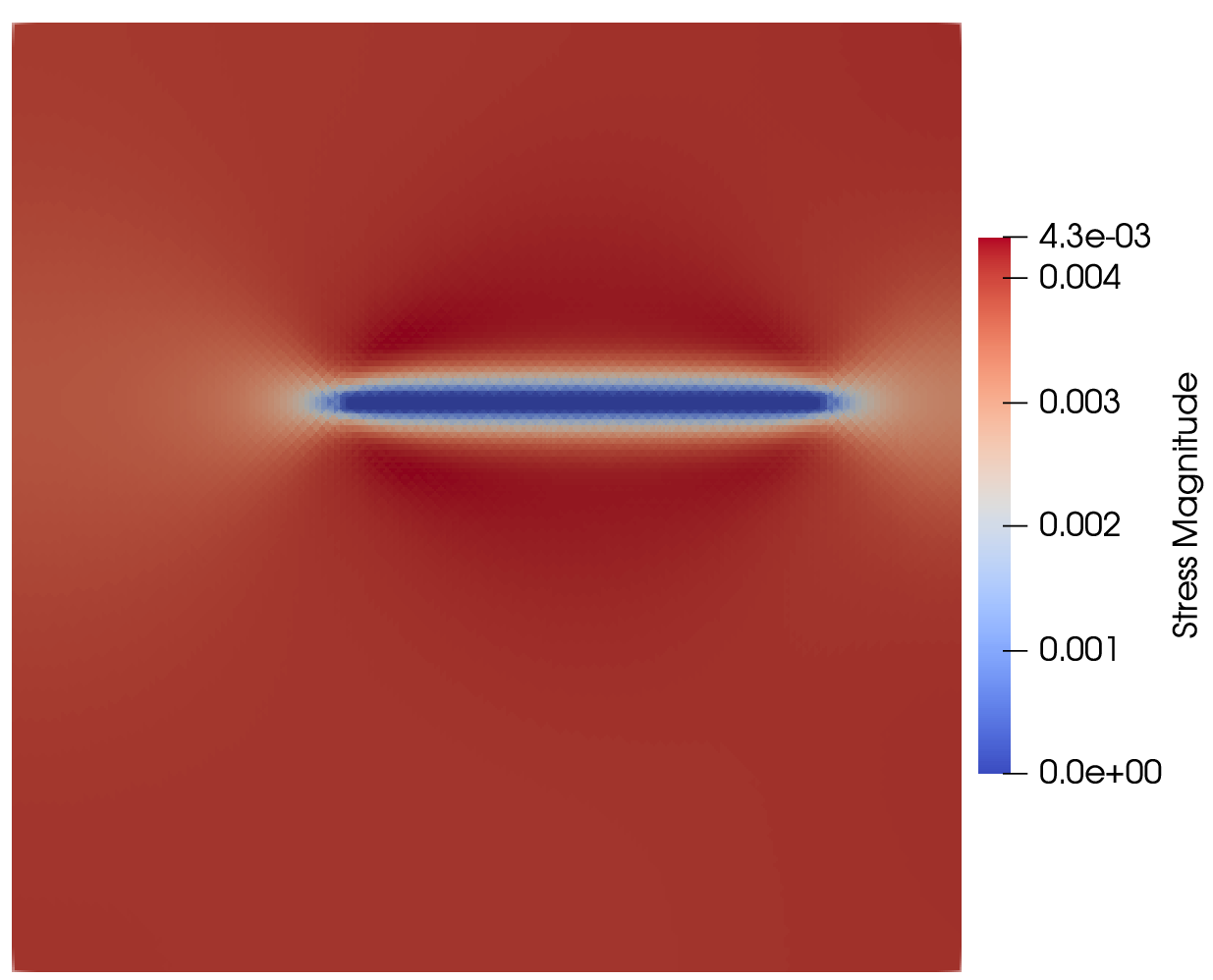}
    }
    \\
    \subfloat[Unphysical crack growth with VolDev splitting.]{
        \includegraphics[width=0.35\textwidth]{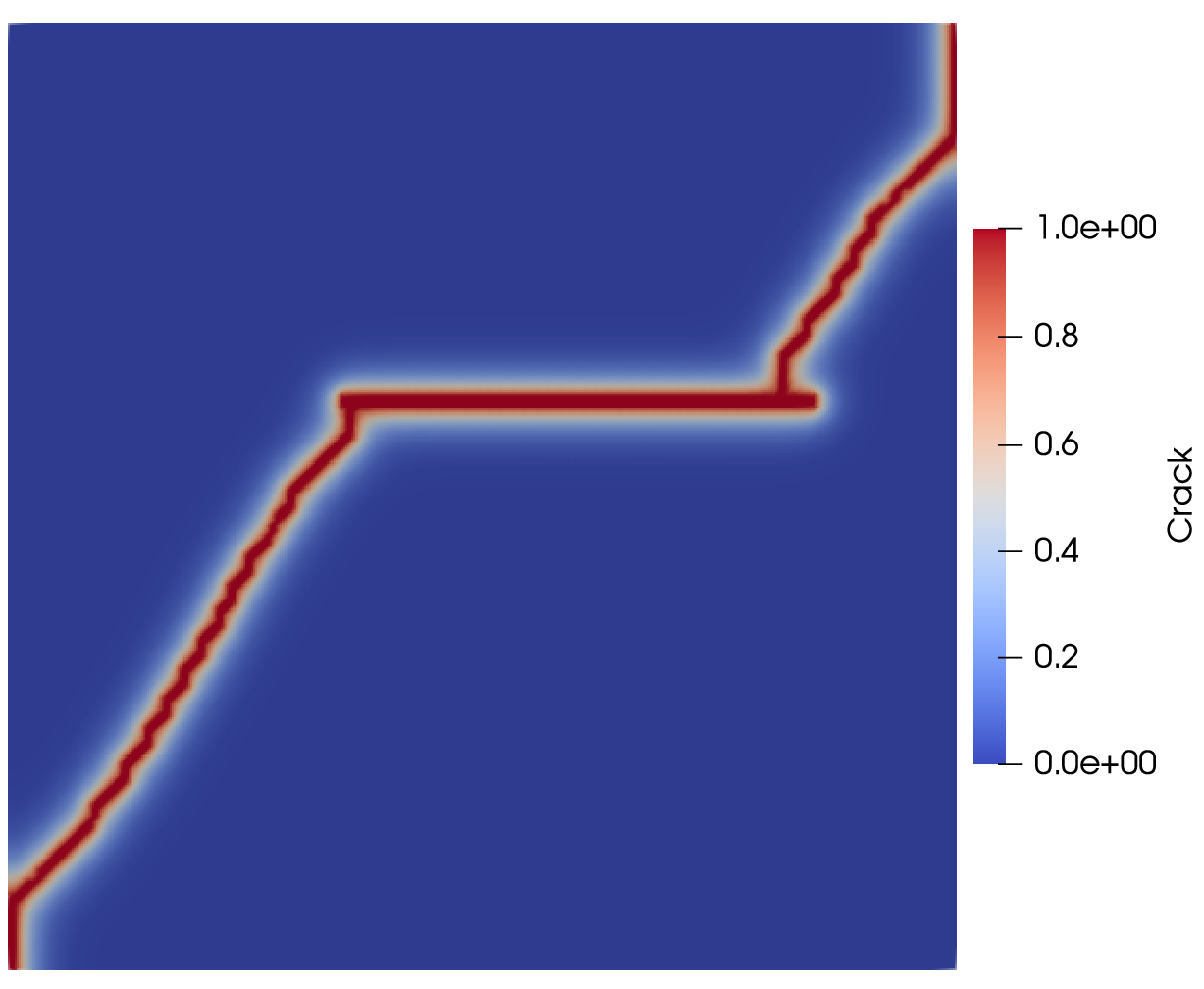}
    }
    \hspace{0.2\textwidth}
    \subfloat[Spurious stress concentrations with VolDev splitting.]{
        \includegraphics[width=0.35\textwidth]{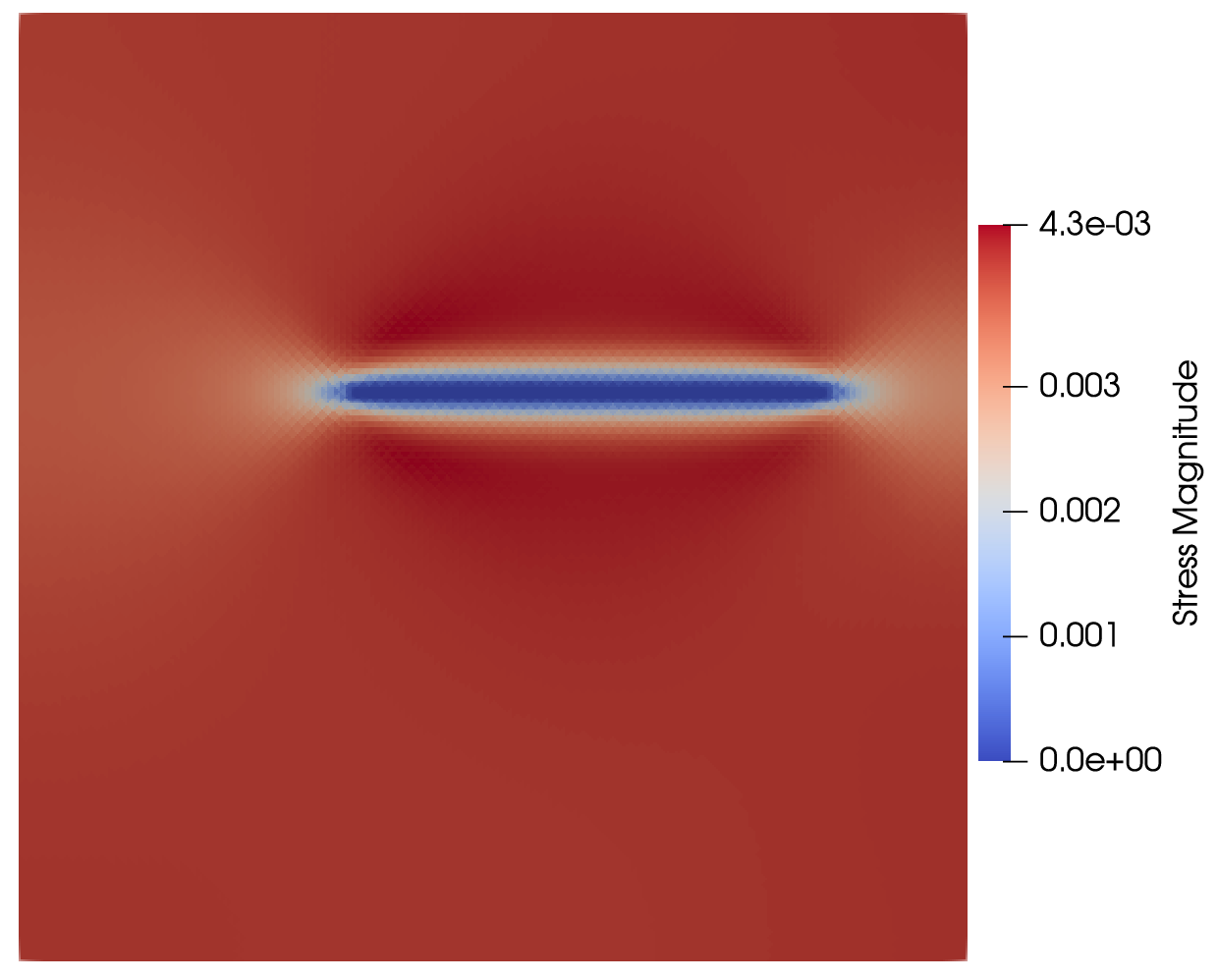}
    }
    \caption{Comparison of crack growth paths and stress fields under compressive crack-parallel stress.}
    \label{fig:T-stress_compare_compression}
\end{figure}

We next examine the effect of compressive loading in Figure \ref{fig:T-stress_compare_compression}, which shows the initial crack, the subsequent crack growth, and the stress distribution at the first load step, for all 3 models; we set $\nu = 0$.
We observe again that our approach predicts essentially a uniform stress distribution, that corresponds to the cracked specimen having the same response as an intact specimen, and there is no crack growth.
Both of these features correspond to physical expectations.

Further, in both the Spectral and VolDev splits, the stresses are significantly non-uniform with spurious stress concentrations at the crack tips that were anticipated by the simple analysis above.
The spurious stress distribution, in turn, drives unphysical crack growth that should simply not occur given that the specimen is under compression.

\paragraph{Effect of Poisson's Ratio.}

In the calculations above, we have set $\nu=0$ to aid the interpretation of the stress fields.
Here, we compare the models with crack-parallel tension and compression, with positive and negative $\nu$.
We examine a horizontal crack in a large rectangular specimen, with the top and bottom traction-free, and the displacement fixed --- to be zero in the vertical direction and nonzero in the horizontal direction --- on the sides.

\begin{figure}[htb!]
    \centering
    \subfloat[Stress field in our model.]{
        \includegraphics[width=0.45\textwidth]{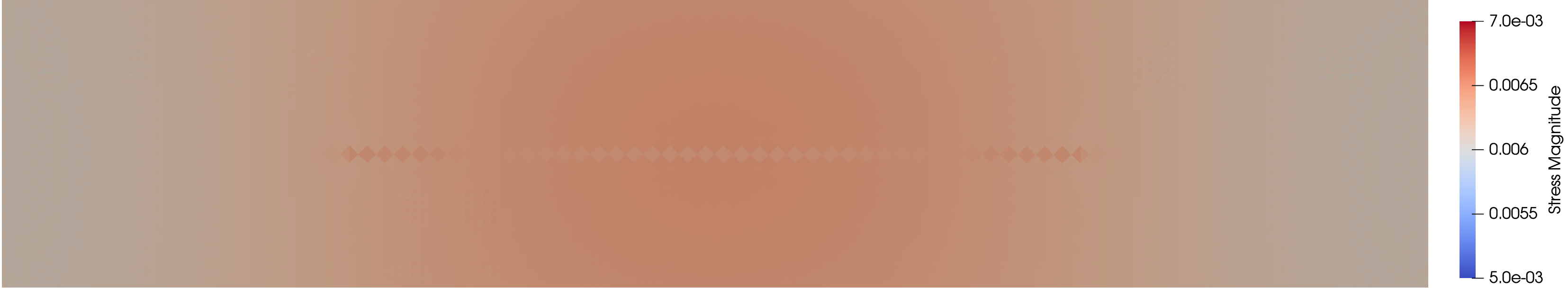}
    }
   \hspace{0.05\textwidth}
    \subfloat[Deformation with our model.]{
        \includegraphics[width=0.45\textwidth]{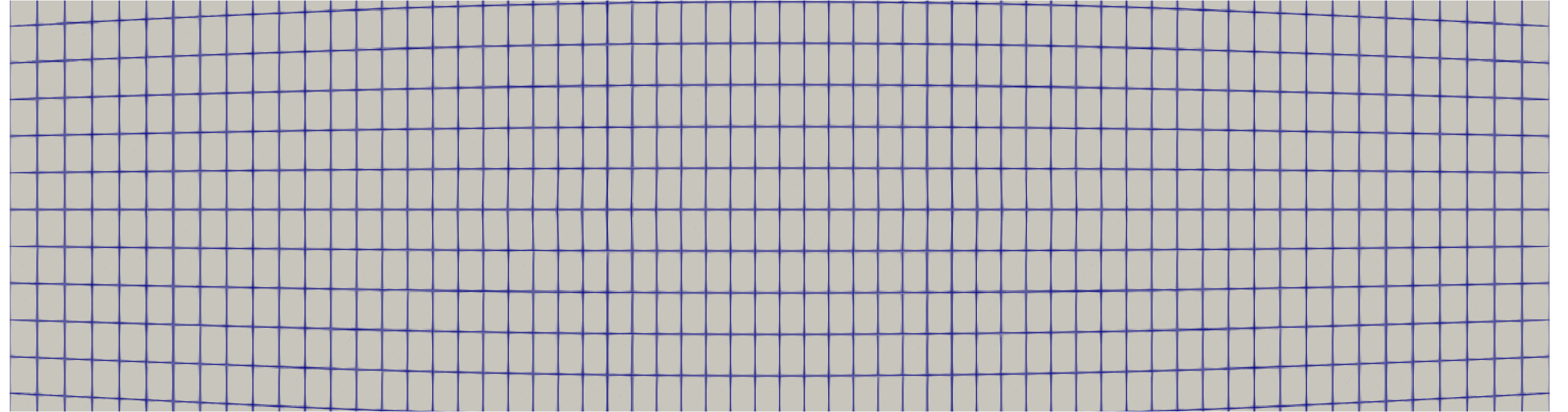}
    }
    \\
    \subfloat[Stress field with Spectral splitting.]{
        \includegraphics[width=0.45\textwidth]{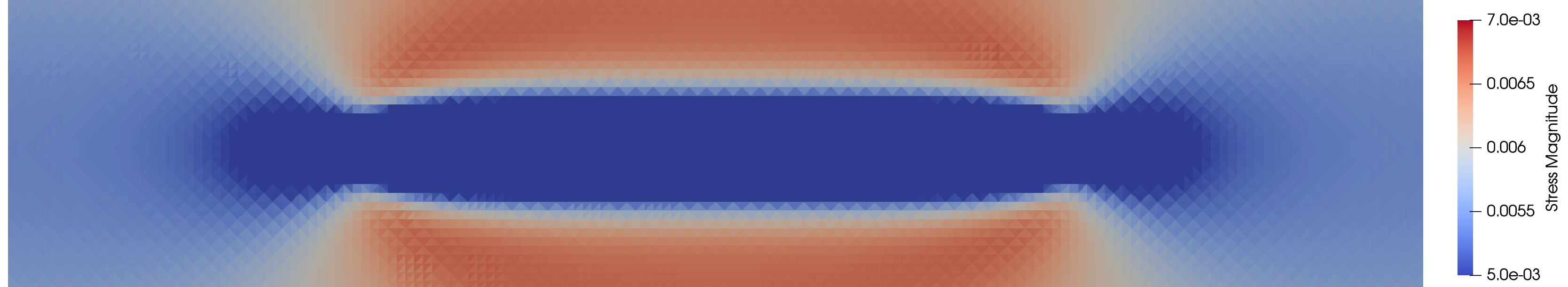}
    }
    \hspace{0.05\textwidth}
    \subfloat[Deformation with Spectral splitting.]{
        \includegraphics[width=0.45\textwidth]{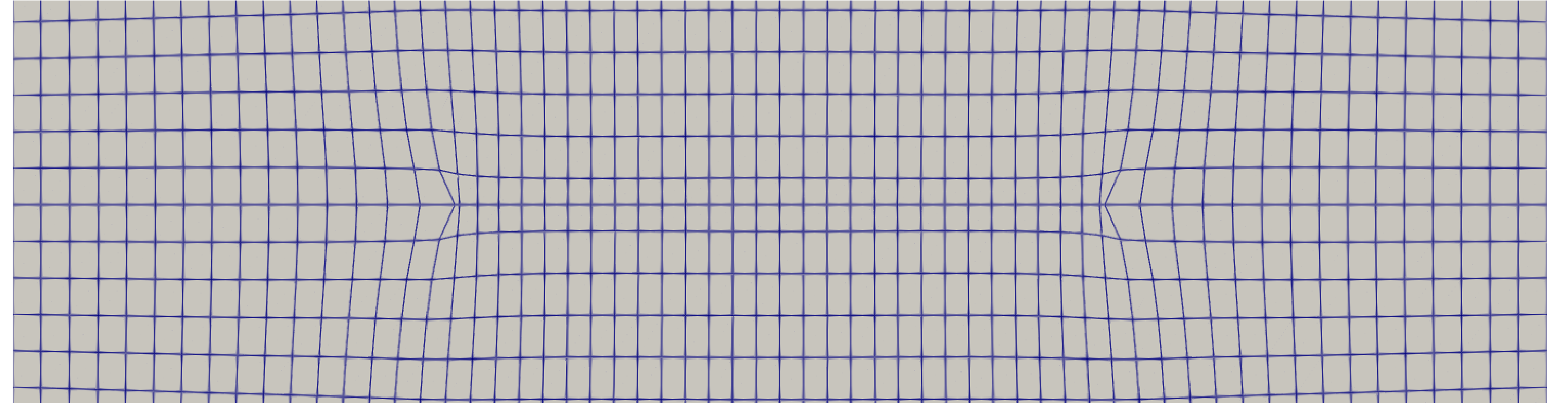}
    }
    \\
    \subfloat[Stress field with VolDev splitting.]{
        \includegraphics[width=0.45\textwidth]{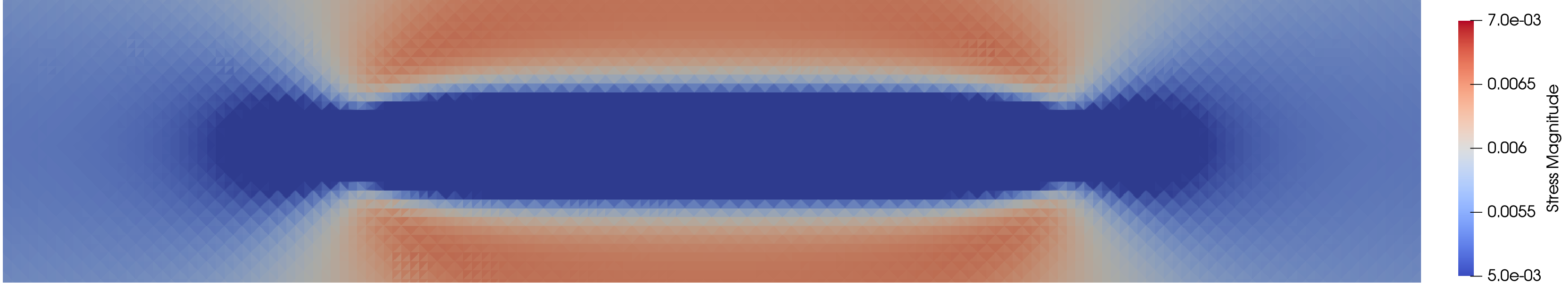}
    }
    \hspace{0.05\textwidth}
    \subfloat[Deformation with VolDev splitting.]{
        \includegraphics[width=0.45\textwidth]{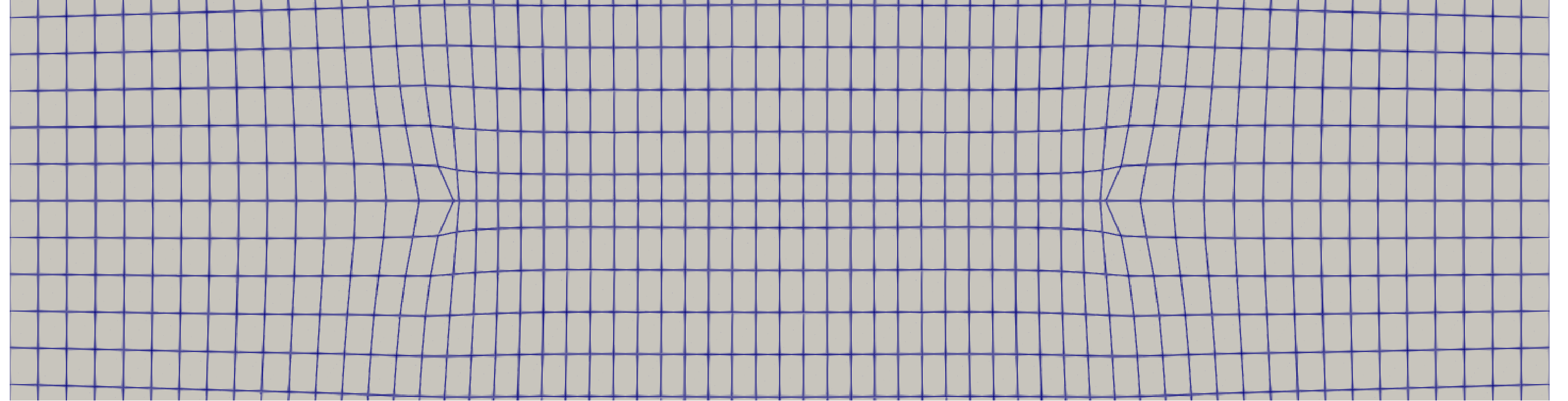}
    }
    \caption{Comparison of stress fields and deformation without crack growth under compressive crack-parallel stress with $\nu>0$. Deformations are magnified $150\times$ for visualization.}
    \label{fig:T-stress_compare_compression_pos_poisson}
\end{figure}

Figure \ref{fig:T-stress_compare_compression_pos_poisson} compares the stresses and deformations under crack-parallel compression with $\nu>0$.
In our model, the stress field is uniform and the deformation corresponds to uniform barreling due to the lateral expansion, as expected physically.
However, in both the Spectral Splitting and VolDev models, the stress is clearly inhomogeneous when it should not be, and the deformation around the crack shows unphysical features.

\begin{figure}[htb!]
    \centering
    \subfloat[Stress field in our model.]{
        \includegraphics[width=0.45\textwidth]{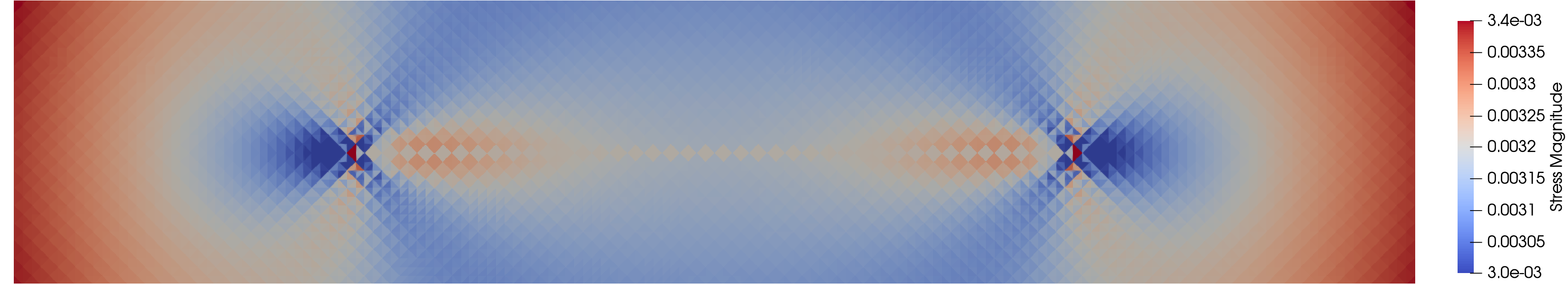}
    }
    \hspace{0.05\textwidth}
    \subfloat[Deformation with our model.]{
        \includegraphics[width=0.45\textwidth]{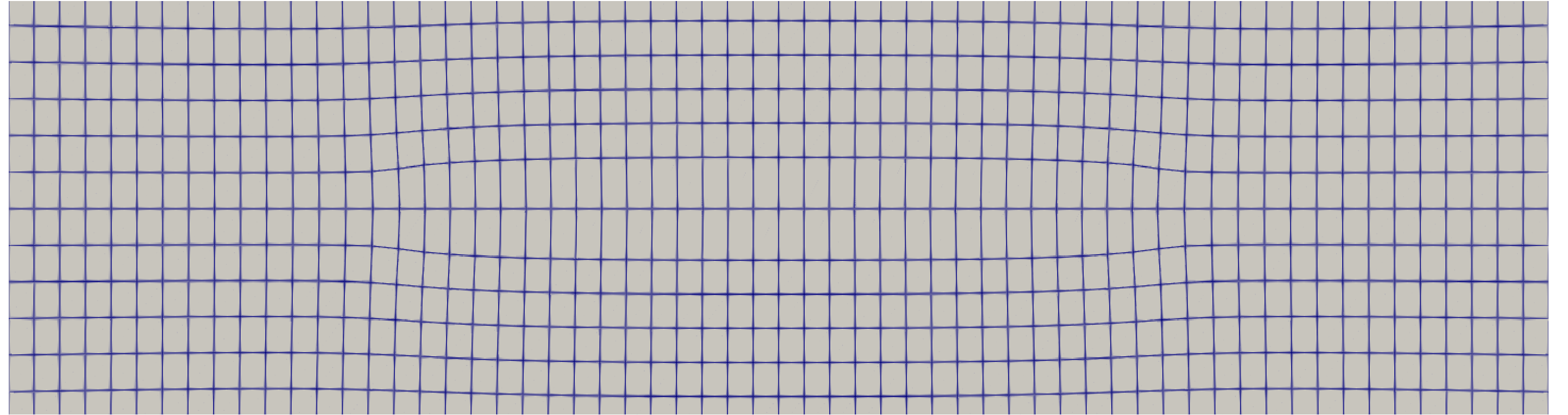}
    }
    \\
    \subfloat[Stress field with Spectral splitting.]{
        \includegraphics[width=0.45\textwidth]{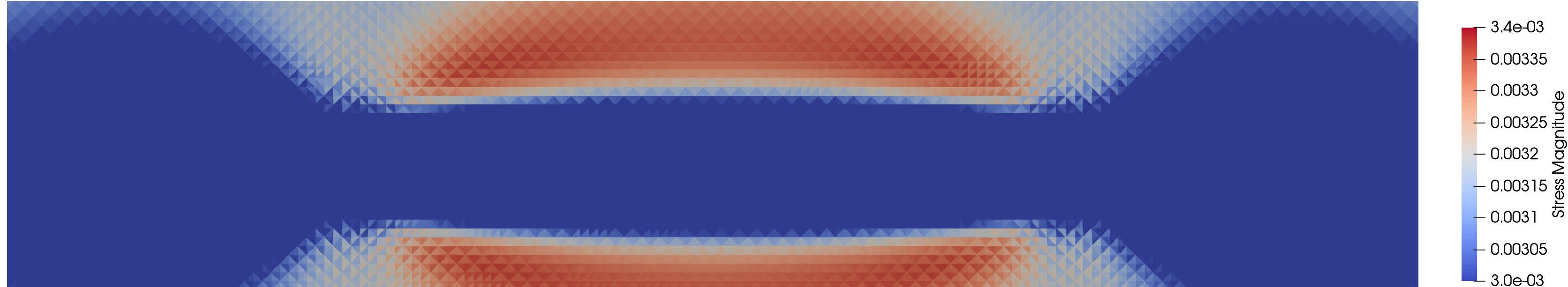}
    }
    \hspace{0.05\textwidth}
    \subfloat[Deformation with Spectral splitting.]{
        \includegraphics[width=0.45\textwidth]{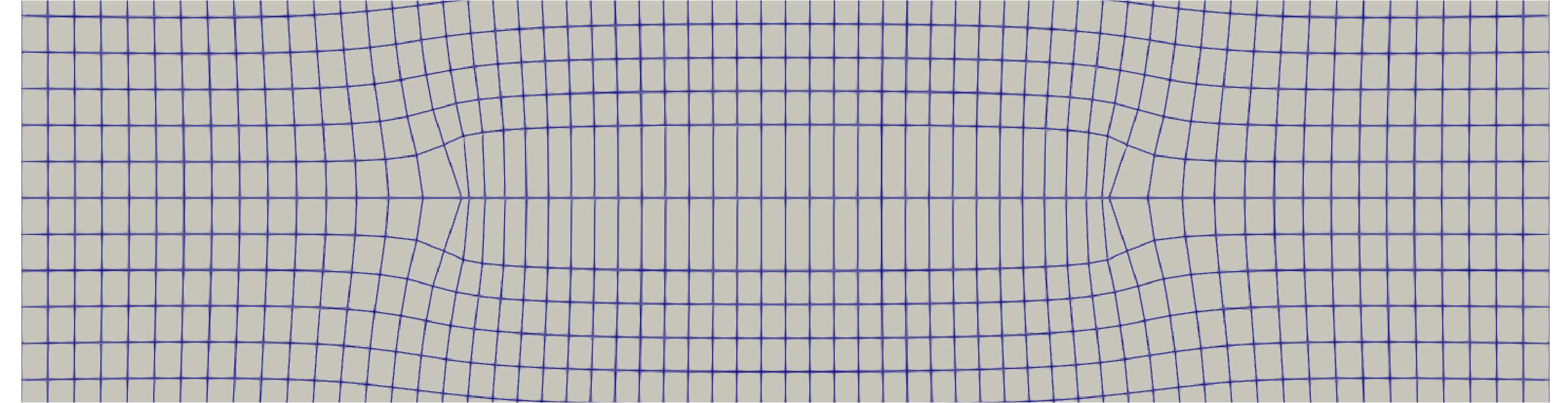}
    }
    \\
    \subfloat[Stress field with VolDev splitting.]{
        \includegraphics[width=0.45\textwidth]{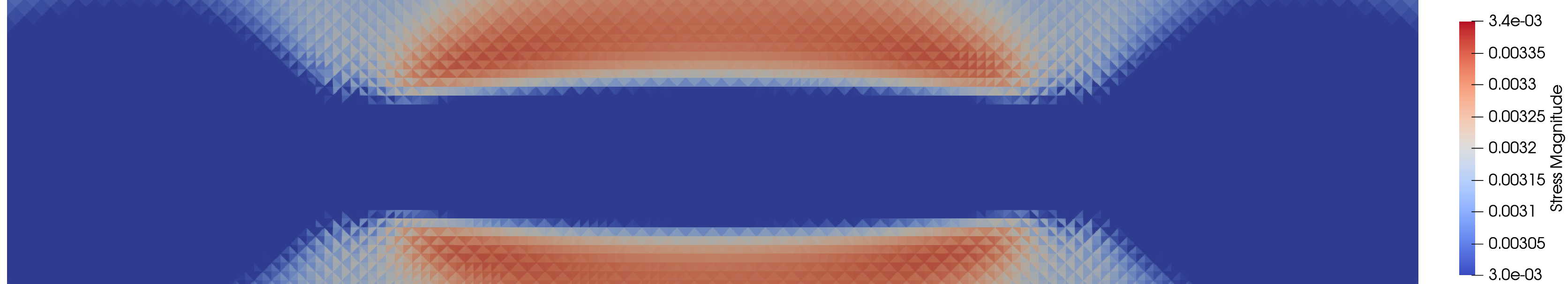}
    }
    \hspace{0.05\textwidth}
    \subfloat[Deformation with VolDev splitting.]{
        \includegraphics[width=0.45\textwidth]{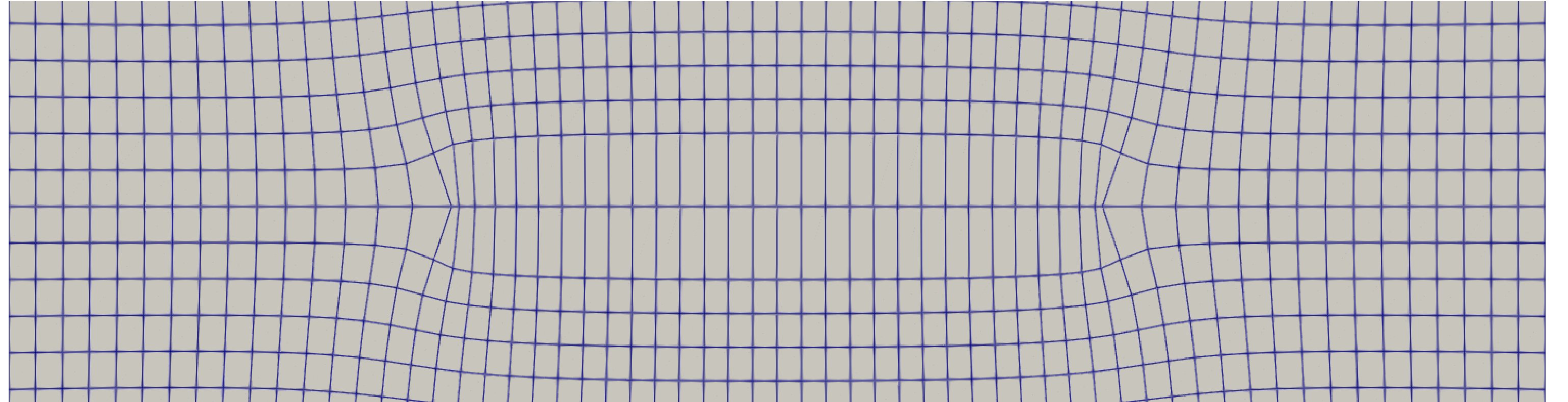}
    }
    \caption{Comparison of stress fields and deformation without crack growth under compressive crack-parallel stress with $\nu<0$. Deformations are magnified $150\times$ for visualization.}
    \label{fig:T-stress_compare_compression_neg_poisson}
\end{figure}

Figure \ref{fig:T-stress_compare_compression_neg_poisson} next compares the models under crack-parallel compression with $\nu<0$.
In our model, as expected, there is lateral contraction that causes the crack faces to lose contact and move apart; the stress field is therefore inhomogeneous with stress concentrations at the crack tips, as is typical.
However, in both the Spectral Splitting and VolDev models, while they appear to capture the lateral contraction, the stress distribution is not at all as expected around crack tips, and the deformation shows unphysical features near the crack tips.

\begin{figure}[htb!]
    \centering
    \subfloat[Stress field in our model.]{
        \includegraphics[width=0.45\textwidth]{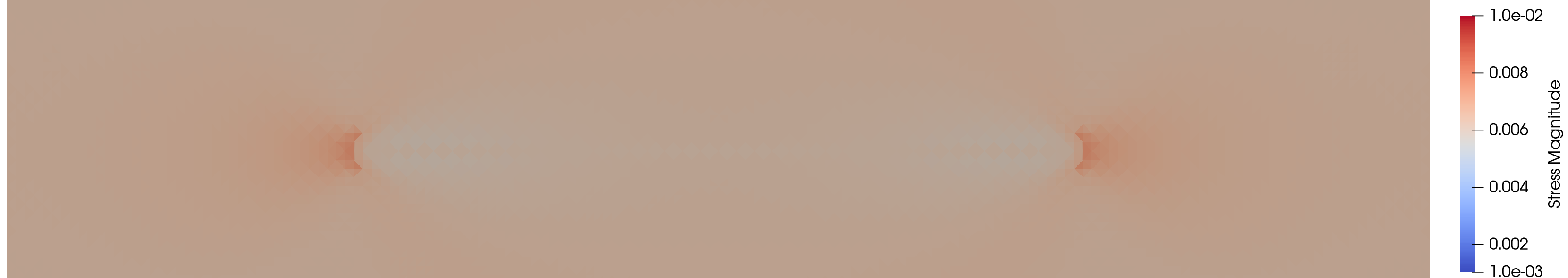}
    }
    \hspace{0.05\textwidth}
    \subfloat[Deformation with our model.]{
        \includegraphics[width=0.45\textwidth]{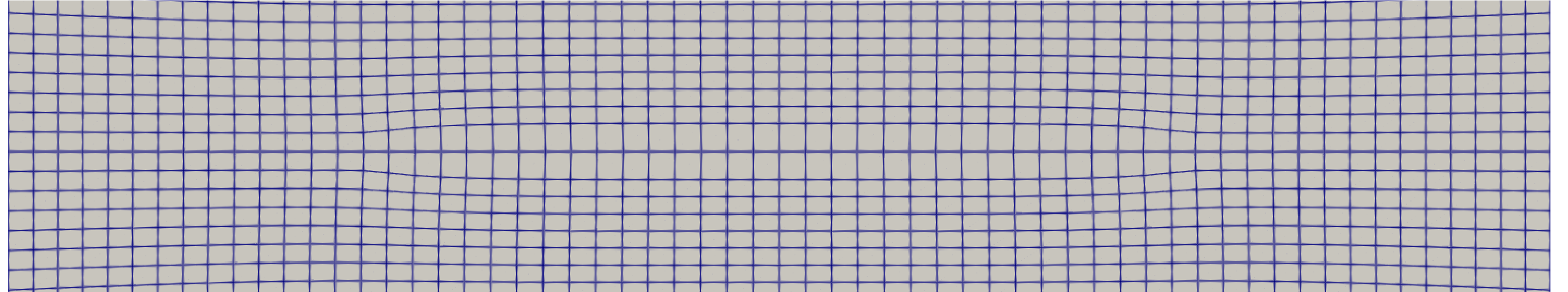}
    }
    \\
    \subfloat[Stress field with Spectral splitting.]{
        \includegraphics[width=0.45\textwidth]{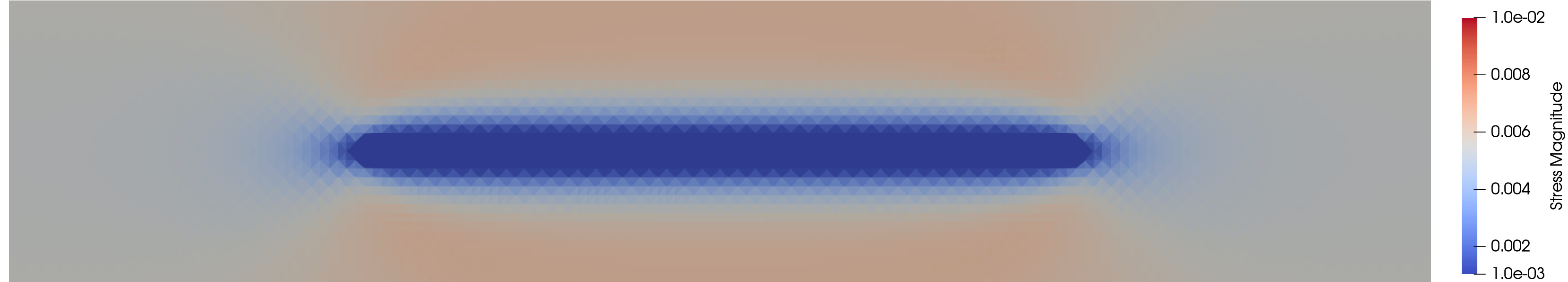}
    }
    \hspace{0.05\textwidth}
    \subfloat[Deformation with Spectral splitting.]{
        \includegraphics[width=0.45\textwidth]{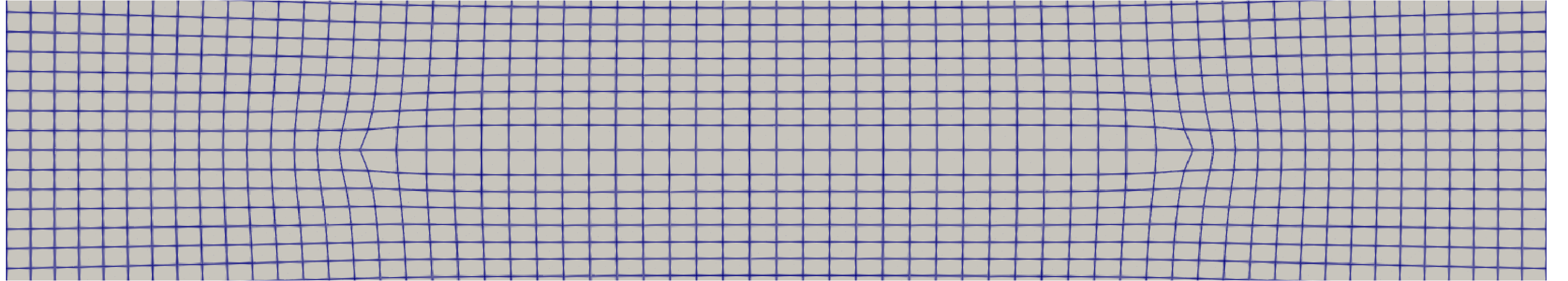}
    }
    \\
    \subfloat[Stress field with VolDev splitting.]{
        \includegraphics[width=0.45\textwidth]{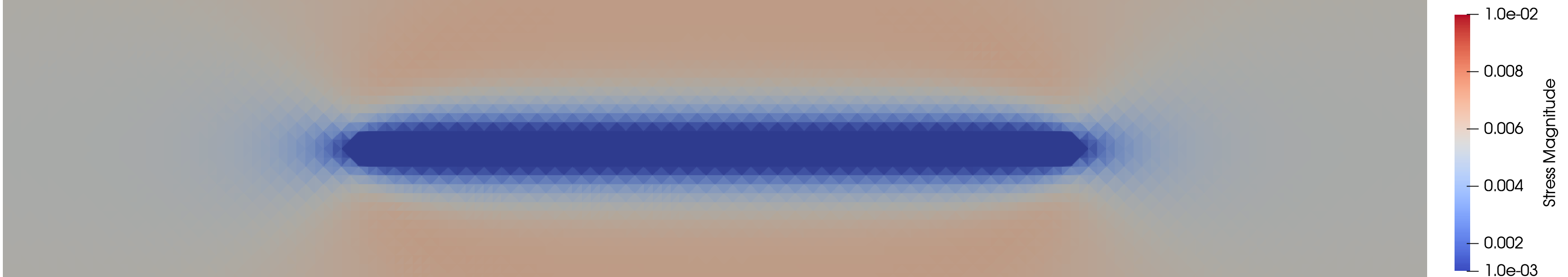}
    }
    \hspace{0.05\textwidth}
    \subfloat[Deformation with VolDev splitting.]{
        \includegraphics[width=0.45\textwidth]{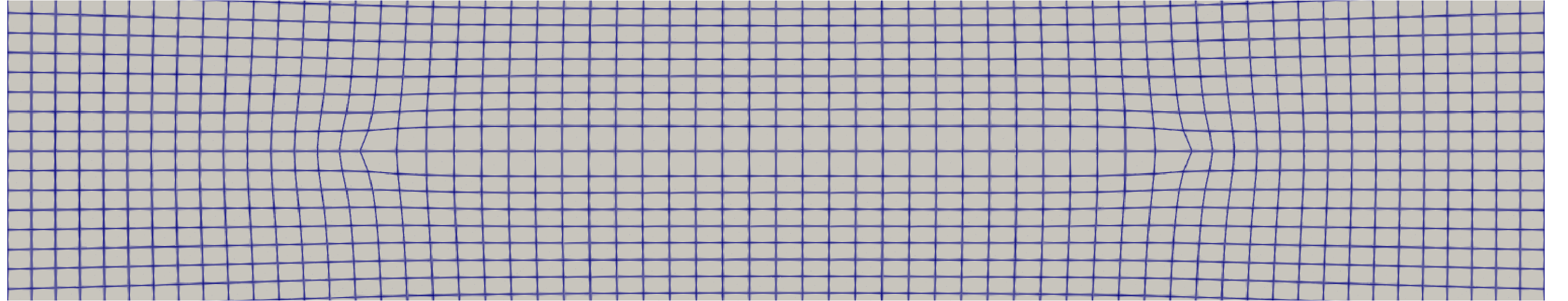}
    }
    \caption{Comparison of stress fields and deformation without crack growth under tensile crack-parallel stress with $\nu>0$. Deformations are magnified $150\times$ for visualization.}
    \label{fig:T-stress_compare_tensile_pos_poisson}
\end{figure}

Figure \ref{fig:T-stress_compare_tensile_pos_poisson} next compares the models under crack-parallel tension with $\nu>0$.
In our model, as expected, there is lateral contraction that causes the crack faces to lose contact and move apart, and the stress field is therefore inhomogeneous with stress concentrations at the crack tips.
However, in both the Spectral Splitting and VolDev models, while they capture the lateral contraction, the stress distribution is not at all as expected around crack tips, and the deformation shows unphysical features near the crack tips.

\begin{figure}[htb!]
    \centering
    \subfloat[Stress field in our model.]{
        \includegraphics[width=0.45\textwidth]{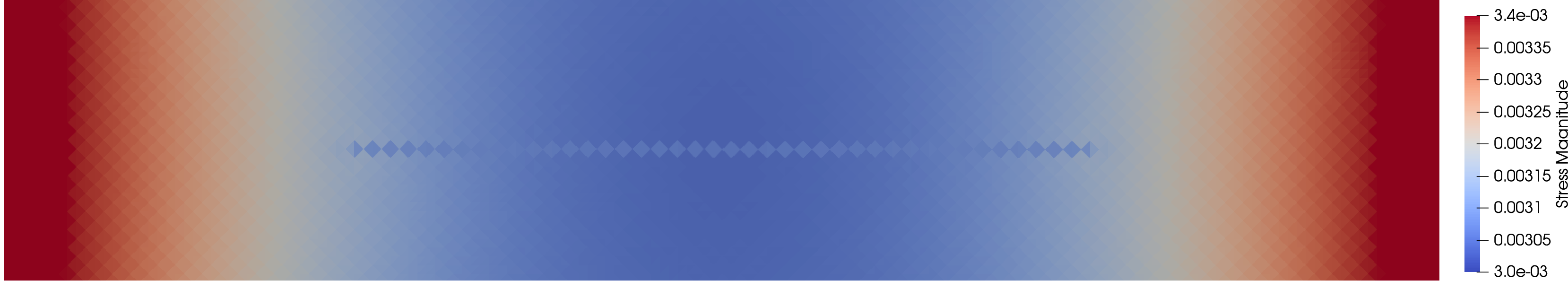}
    }
    \hspace{0.05\textwidth}
    \subfloat[Deformation with our model.]{
        \includegraphics[width=0.45\textwidth]{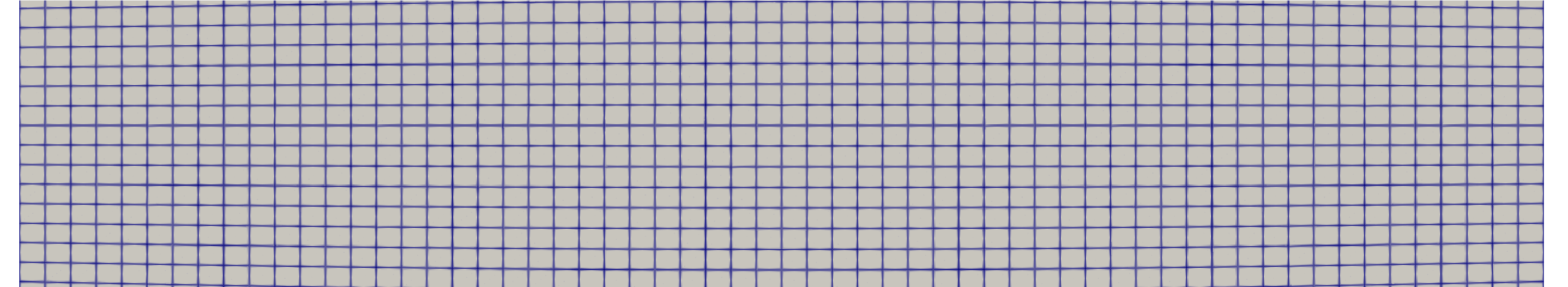}
    }
    \\
    \subfloat[Stress field with Spectral splitting.]{
        \includegraphics[width=0.45\textwidth]{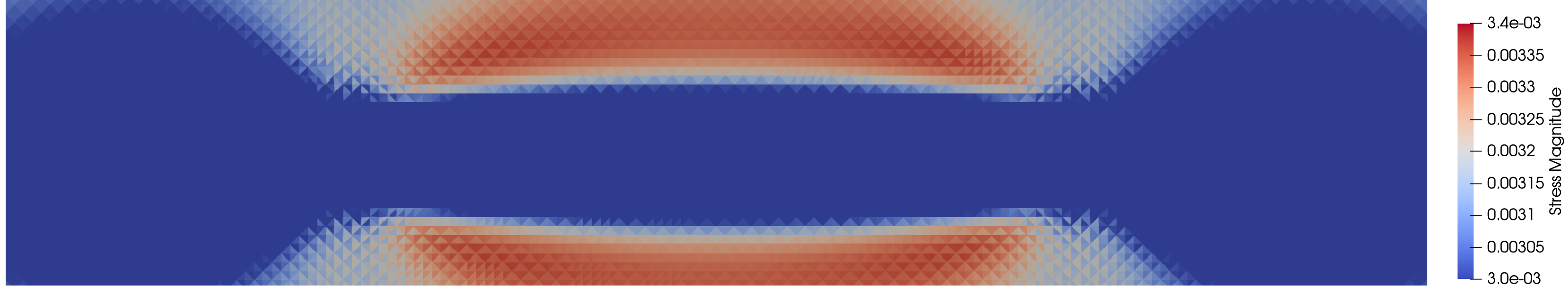}
    }
    \hspace{0.05\textwidth}
    \subfloat[Deformation with Spectral splitting.]{
        \includegraphics[width=0.45\textwidth]{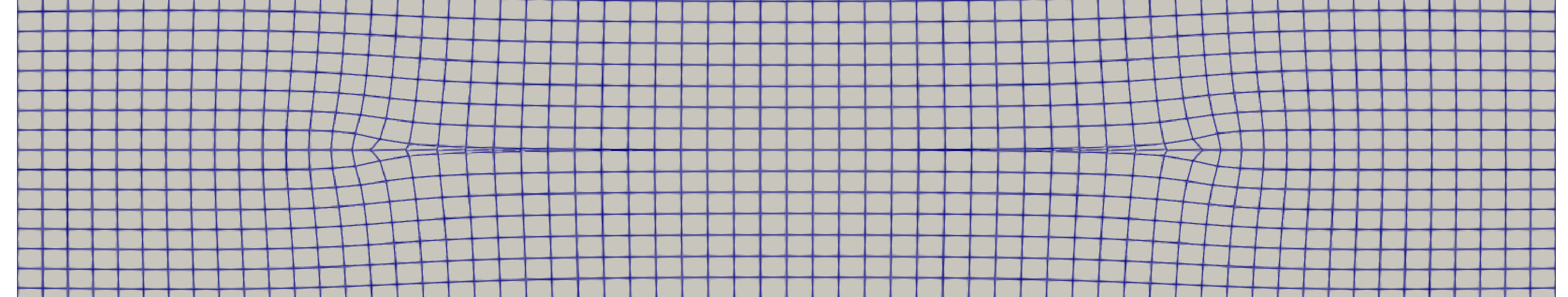}
    }
    \\
    \subfloat[Stress field with VolDev splitting.]{
        \includegraphics[width=0.45\textwidth]{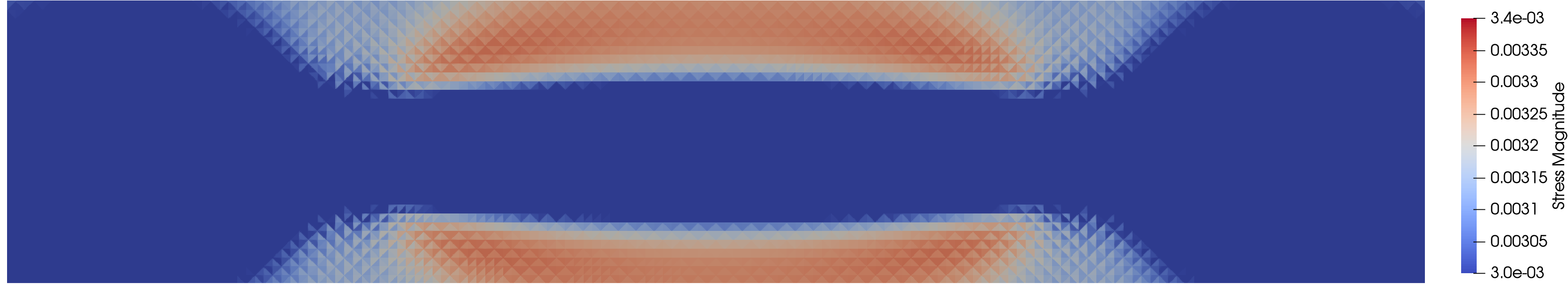}
    }
    \hspace{0.05\textwidth}
    \subfloat[Deformation with VolDev splitting.]{
        \includegraphics[width=0.45\textwidth]{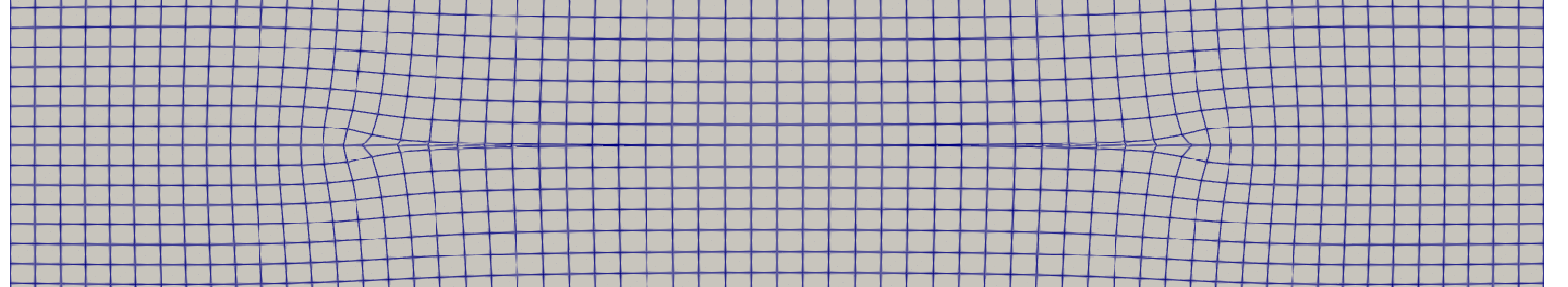}
    }
    \caption{Comparison of stress fields and deformation without crack growth under tensile crack-parallel stress with $\nu<0$. Deformations are magnified $150\times$ for visualization.}
    \label{fig:T-stress_compare_tensile_neg_poisson}
\end{figure}

Finally, Figure \ref{fig:T-stress_compare_tensile_neg_poisson} compares the models under crack-parallel tension with $\nu<0$.
In our model, as expected, there is lateral expansion and the overall deformation is uniform as the crack faces remain in contact; the stress field is similarly homogeneous around the crack.
However, in both the Spectral Splitting and VolDev models, the deformation is completely unphysical: the material within the damaged region shrinks laterally, corresponding to the crack faces inter-penetrating.
Further, the stress distribution is not at all as expected around crack tips.

\section{Concluding Remarks}

We have shown that the widely-used Spectral and VolDev splitting approaches lead to unphysical crack growth and spurious stress concentrations in the important settings of crack-parallel stresses.
We have further shown that accounting for crack-face contact through the crack-response model proposed in \cite{hakimzadeh2022phase} provides the correct predictions in these settings, suggesting the use of this model in situations where crack-parallel stresses play an important role \cite{bavzant2022critical}.

We have idealized the crack as having no friction between the crack faces for simplicity.
For future studies aiming to infuse more realistic physics into the model, it is important to further incorporate friction in the model \cite{fei2020phase-cmame,fei2020phase-ijnme}. 
Another important area of application is to couple the model to poromechanics, e.g., by adapting \cite{karimi2022energetic,karimi2023high,chua2024deformation,sun2021poro,clayton2020constitutive,duddu2020non}.
A further important area is to combine the crack-response model with crack nucleation models, following \cite{agrawal-dayal-2015a,agrawal-dayal-2015b,chua2022phase,chua2024interplay,agrawal2017dependence}.
For all of these future directions, it is essential to predict the correct stress fields that in turn provide the correct crack growth driving force.

Our focus in this paper has been on the small-strain regime for simplicity.
However, the crack-response model of \cite{hakimzadeh2022phase} is valid in the large deformation setting and can be readily incorporated in such models, e.g., \cite{clayton2014geometrically,clayton2015nonlinear,clayton2021nonlinear,mousavi2024evaluating,mousavi2025chain}.


\paragraph*{Software and Data Availability.}

The code implementing all three approaches for the calculations in this paper is available at \\ 
\url{https://github.com/maryhzd/Phase-Field-Fracture-T-stress}.  

\paragraph*{Acknowledgments.}

We thank Meenu Krishnan for useful discussions. We acknowledge financial support from NSF (2108784, 2012259, 2342349), ARO (MURI W911NF-24-2-0184), AEI (PID2021-124195NB-C32, PCI2024-155023-2, CEX2023-001347-S funded by MICIU/AEI/10.13039/501100011033) and ERC (834728); and NSF for XSEDE computing resources provided by Pittsburgh Supercomputing Center.


\end{document}